\documentclass[aps,prc,superscriptaddress,twocolumn,nofootinbib,amsmath,amssymb,showpacs,showkeys,floatfix,preprintnumbers]{revtex4-1}

\usepackage{pifont}
\usepackage{graphicx}
\usepackage{color}
\usepackage{amsmath}
\usepackage{amssymb}
\usepackage{lipsum}
\usepackage{setspace}

\usepackage[caption=false]{subfig}
\usepackage{multirow}

\usepackage{longtable}
\usepackage[percent]{overpic}

\graphicspath{ {figures/} }

\newcommand{\about}{\raise.17ex\hbox{$\scriptstyle\sim$}}
\newcommand{\etapr}{\mbox{$\eta^\prime$}(958)}

\newcommand{\xx}{\mbox{$x$}}
\newcommand{\fx}{\mbox{$f_1(1285)$}}
\newcommand{\etax}{\mbox{$\eta(1295)$}}

\newcommand{\etapipi}{\mbox{$\eta\pi^+\pi^-$}}
\newcommand{\rhogamma}{\mbox{$\gamma\rho^0$}}
\newcommand{\kkpi}{\mbox{$K^\pm K^0 \pi^\mp$}}

\def \clas {{\textsc{CLAS}}}

\def \mmp  {{missing mass off the proton}}

\def \MMp  {{Missing mass off the proton}}
\def \mmunit {{${\rm  MeV}/c^2$}}              

\begin{document}



\title{Photoproduction of the $f_1(1285)$ Meson}

\newcommand*{\CMU}{Carnegie Mellon University, Pittsburgh, Pennsylvania 15213, USA}
\newcommand*{\CMUindex}{1}
\affiliation{\CMU}
\newcommand*{\ANL}{Argonne National Laboratory, Argonne, Illinois 60439, USA}
\newcommand*{\ANLindex}{2}
\affiliation{\ANL}
\newcommand*{\ASU}{Arizona State University, Tempe, Arizona 85287, USA}
\newcommand*{\ASUindex}{3}
\affiliation{\ASU}
\newcommand*{\CSUDH}{California State University, Dominguez Hills, Carson, CA 90747, USA}
\newcommand*{\CSUDHindex}{4}
\affiliation{\CSUDH}
\newcommand*{\CANISIUS}{Canisius College, Buffalo, NY, USA}
\newcommand*{\CANISIUSindex}{5}
\affiliation{\CANISIUS}
\newcommand*{\CUA}{Catholic University of America, Washington, D.C. 20064, USA}
\newcommand*{\CUAindex}{6}
\affiliation{\CUA}
\newcommand*{\SACLAY}{CEA, Centre de Saclay, Irfu/Service de Physique Nucl\'eaire, 91191 Gif-sur-Yvette, France}
\newcommand*{\SACLAYindex}{7}
\affiliation{\SACLAY}
\newcommand*{\UCONN}{University of Connecticut, Storrs, Connecticut 06269, USA}
\newcommand*{\UCONNindex}{8}
\affiliation{\UCONN}
\newcommand*{\FU}{Fairfield University, Fairfield, Connecticut 06824, USA}
\newcommand*{\FUindex}{9}
\affiliation{\FU}
\newcommand*{\FIU}{Florida International University, Miami, Florida 33199, USA}
\newcommand*{\FIUindex}{10}
\affiliation{\FIU}
\newcommand*{\FSU}{Florida State University, Tallahassee, Florida 32306, USA}
\newcommand*{\FSUindex}{11}
\affiliation{\FSU}
\newcommand*{\Genova}{Universit$\grave{a}$ di Genova, 16146 Genova, Italy}
\newcommand*{\Genovaindex}{12}
\affiliation{\Genova}
\newcommand*{\GWUI}{The George Washington University, Washington, DC 20052, USA}
\newcommand*{\GWUIindex}{13}
\affiliation{\GWUI}
\newcommand*{\ISU}{Idaho State University, Pocatello, Idaho 83209, USA}
\newcommand*{\ISUindex}{14}
\affiliation{\ISU}
\newcommand*{\INFNFE}{INFN, Sezione di Ferrara, 44100 Ferrara, Italy}
\newcommand*{\INFNFEindex}{15}
\affiliation{\INFNFE}
\newcommand*{\INFNFR}{INFN, Laboratori Nazionali di Frascati, 00044 Frascati, Italy}
\newcommand*{\INFNFRindex}{16}
\affiliation{\INFNFR}
\newcommand*{\INFNGE}{INFN, Sezione di Genova, 16146 Genova, Italy}
\newcommand*{\INFNGEindex}{17}
\affiliation{\INFNGE}
\newcommand*{\INFNRO}{INFN, Sezione di Roma Tor Vergata, 00133 Rome, Italy}
\newcommand*{\INFNROindex}{18}
\affiliation{\INFNRO}
\newcommand*{\INFNTUR}{INFN, Sezione di Torino, 10125 Torino, Italy}
\newcommand*{\INFNTURindex}{19}
\affiliation{\INFNTUR}
\newcommand*{\ORSAY}{Institut de Physique Nucl\'eaire, CNRS/IN2P3 and Universit\'e Paris Sud, Orsay, France}
\newcommand*{\ORSAYindex}{20}
\affiliation{\ORSAY}
\newcommand*{\ITEP}{Institute of Theoretical and Experimental Physics, Moscow, 117259, Russia}
\newcommand*{\ITEPindex}{21}
\affiliation{\ITEP}
\newcommand*{\JMU}{James Madison University, Harrisonburg, Virginia 22807, USA}
\newcommand*{\JMUindex}{22}
\affiliation{\JMU}
\newcommand*{\KNU}{Kyungpook National University, Daegu 702-701, Republic of Korea}
\newcommand*{\KNUindex}{23}
\affiliation{\KNU}
\newcommand*{\MISS}{Mississippi State University, Mississippi State, MS 39762, USA}
\newcommand*{\MISSindex}{24}
\affiliation{\MISS}
\newcommand*{\UNH}{University of New Hampshire, Durham, New Hampshire 03824, USA}
\newcommand*{\UNHindex}{25}
\affiliation{\UNH}
\newcommand*{\NSU}{Norfolk State University, Norfolk, Virginia 23504, USA}
\newcommand*{\NSUindex}{26}
\affiliation{\NSU}
\newcommand*{\OHIOU}{Ohio University, Athens, Ohio  45701, USA}
\newcommand*{\OHIOUindex}{27}
\affiliation{\OHIOU}
\newcommand*{\ODU}{Old Dominion University, Norfolk, Virginia 23529, USA}
\newcommand*{\ODUindex}{28}
\affiliation{\ODU}
\newcommand*{\RPI}{Rensselaer Polytechnic Institute, Troy, New York 12180, USA}
\newcommand*{\RPIindex}{29}
\affiliation{\RPI}
\newcommand*{\URICH}{University of Richmond, Richmond, Virginia 23173, USA}
\newcommand*{\URICHindex}{30}
\affiliation{\URICH}
\newcommand*{\ROMAII}{Universita' di Roma Tor Vergata, 00133 Rome Italy}
\newcommand*{\ROMAIIindex}{31}
\affiliation{\ROMAII}
\newcommand*{\MSU}{Skobeltsyn Institute of Nuclear Physics, Lomonosov Moscow State University, 119234 Moscow, Russia}
\newcommand*{\MSUindex}{32}
\affiliation{\MSU}
\newcommand*{\SCAROLINA}{University of South Carolina, Columbia, South Carolina 29208, USA}
\newcommand*{\SCAROLINAindex}{33}
\affiliation{\SCAROLINA}
\newcommand*{\TEMPLE}{Temple University,  Philadelphia, PA 19122, USA }
\newcommand*{\TEMPLEindex}{34}
\affiliation{\TEMPLE}
\newcommand*{\JLAB}{Thomas Jefferson National Accelerator Facility, Newport News, Virginia 23606, USA}
\newcommand*{\JLABindex}{35}
\affiliation{\JLAB}
\newcommand*{\UTFSM}{Universidad T\'{e}cnica Federico Santa Mar\'{i}a, Casilla 110-V Valpara\'{i}so, Chile}
\newcommand*{\UTFSMindex}{36}
\affiliation{\UTFSM}
\newcommand*{\EDINBURGH}{Edinburgh University, Edinburgh EH9 3JZ, United Kingdom}
\newcommand*{\EDINBURGHindex}{37}
\affiliation{\EDINBURGH}
\newcommand*{\GLASGOW}{University of Glasgow, Glasgow G12 8QQ, United Kingdom}
\newcommand*{\GLASGOWindex}{38}
\affiliation{\GLASGOW}
\newcommand*{\VIRGINIA}{University of Virginia, Charlottesville, Virginia 22901, USA}
\newcommand*{\VIRGINIAindex}{39}
\affiliation{\VIRGINIA}
\newcommand*{\WM}{College of William and Mary, Williamsburg, Virginia 23187, USA}
\newcommand*{\WMindex}{40}
\affiliation{\WM}
\newcommand*{\YEREVAN}{Yerevan Physics Institute, 375036 Yerevan, Armenia}
\newcommand*{\YEREVANindex}{41}
\affiliation{\YEREVAN}
 
\newcommand*{\NOWJLAB}{Thomas Jefferson National Accelerator Facility, Newport News, Virginia 23606}
\newcommand*{\NOWODU}{Old Dominion University, Norfolk, Virginia 23529}
\newcommand*{\NOWINFNGE}{INFN, Sezione di Genova, 16146 Genova, Italy}
\newcommand*{\NOWEDINBURGH}{Edinburgh University, Edinburgh EH9 3JZ, United Kingdom}

\author{R. Dickson}
\altaffiliation[Current address:  UPMC Enterprises, Pittsburgh, PA 15044]{}
\affiliation{\CMU}
\email[current address ]{UMPC, Pittsburgh, PA}
\author{R. A. Schumacher}
\altaffiliation[Contact: schumacher@cmu.edu]{}
\affiliation{\CMU}

\author {K.P. ~Adhikari} 
\affiliation{\MISS}
\author {Z.~Akbar} 
\affiliation{\FSU}
\author {M.J.~Amaryan} 
\affiliation{\ODU}
\author {S. ~Anefalos~Pereira} 
\affiliation{\INFNFR}
\author {R.A.~Badui} 
\affiliation{\FIU}
\author {J.~Ball} 
\affiliation{\SACLAY}
\author {M.~Battaglieri} 
\affiliation{\INFNGE}
\author {V.~Batourine} 
\affiliation{\JLAB}
\affiliation{\KNU}
% ...

\collaboration{CLAS Collaboration}
\noaffiliation



\date{\today}

\begin{abstract}
 The \fx\ meson with mass $1281.0 \pm 0.8$~MeV/$c^2$ and width  $18.4  \pm 1.4$~MeV (FWHM) was measured for the first time in photoproduction from a proton target using CLAS at  Jefferson Lab. Differential cross sections were obtained via the $\eta\pi^{+}\pi^{-}$, $K^+\bar{K}^0\pi^-$, and $K^-K^0\pi^+$ decay channels from threshold up to a center-of-mass  energy of 2.8~GeV. The mass, width, and an amplitude analysis of the  $\eta\pi^{+}\pi^{-}$ final-state Dalitz distribution are consistent with the  axial-vector $J^P=1^+$ \fx\ identity, rather than the pseudoscalar $0^-$ \etax.  The production mechanism is more consistent with $s$-channel decay of a high-mass $N^*$ state, and not with $t$-channel meson exchange.  Decays to $\eta\pi\pi$ go dominantly via the intermediate  $a_0^\pm(980)\pi^\mp$ states, with the branching ratio $\Gamma(a_0\pi \text{\small{ (no }} \bar{K} K\text{\small{)}}) / \Gamma(\eta\pi\pi \text{\small{ (all)}}) = 0.74\pm0.09$.  The branching ratios $\Gamma(K \bar{K} \pi)/\Gamma(\eta\pi\pi) = 0.216\pm0.033$ and  $\Gamma(\gamma\rho^0)/\Gamma(\eta\pi\pi) = 0.047\pm0.018$ were also obtained.  The first  is in agreement with previous data for the \fx, while the latter is lower than the world average.  
\end{abstract}


\maketitle


\section{Introduction}
\label{sec:intro}


The \fx\ fits well into the Quark Model as a member of the $^3P_1$ axial-vector nonet, the isoscalar flavor-mixing partner of the $f_1(1420)$.
The \fx\ meson was discovered in $p\bar{p}$ annihilation independently at BNL~\cite{Miller:1965zza} and at CERN~\cite{d'Andlau:1965zz} in 1965.  Both experiments observed a resonance decaying to $K\bar{K}\pi$, with the quantum numbers $I^G(J^{PC}) = 0^+(1^{++})$ that were definitively confirmed in Ref.~\cite{Debilly:1980bc}.  Two more recent experiments have made very clean measurements of the \fx\ in $pp$ central production and in  $\gamma\gamma$ collisions.  Experiment WA102 observed the \fx\ and $f_1(1420)$ mesons decaying to $\eta\pi\pi$, \rhogamma~\cite{Barberis:1998by}, four pion~\cite{Barberis:1997ve,Barberis:1999wn}, and $K \bar{K}\pi$~\cite{Barberis:1997vf} final states.  They found no evidence for $0^+(0^{-+})$ $\eta$-like pseudoscalar states in this mass region, in agreement with earlier central production experiments WA76~\cite{Armstrong:1991rg} at CERN and E690~\cite{Sosa:1997qm} at Fermilab.  The suppression of $0^{-+}$ production in central production allowed WA102 to measure \fx\ branching fractions in the major decay channels with good accuracy without concern about possible \etax\ contamination.

The L3 Collaboration at CERN observed the \fx\ in virtual two-photon collision events $e^+e^- \to e^+e^-\gamma_v\gamma_v \to e^+e^-\eta\pi^+\pi^-$. With increasing virtuality of the photons ($Q^2 > 0$), the relative production of a spin-0 state to a spin-1 state diminishes ~\cite{PhysRev.77.242}.  The experiment therefore separated the pseudoscalar and axial-vector contributions to the data by binning their spectra in total transverse momentum, $P_T$, which approximates $Q^2$. They observed the \fx\ decaying to both $\eta\pi\pi$ and $K_S^0 K^\pm \pi^\mp$, and set an upper-limit on two-photon production of the \etax, $\Gamma_{\gamma\gamma}(\etax) \times BR(\etax \to \eta\pi\pi) < 66\;{\rm eV}$~\cite{Acciarri:2000ev}.  The analysis of \fx\ decays to $\eta\pi^+\pi^-$  found the branching ratio  $\Gamma(\fx \to a_0\pi)/\Gamma(\fx \to \eta\pi\pi)$ consistent with 100$\%$  and with a lower limit of 69\%  at  confidence level of 95$\%$~\cite{Achard:2001uu}.  This is notable since the present results are consistent with an intermediate value.


A pseudoscalar meson of nearly the same mass, the \etax, was first observed in partial-wave analysis of $\pi^- p \to \eta \pi^+ \pi^- n$ data obtained at the Argonne National Laboratory ZGS~\cite{Stanton:1979ya}.  Another observation came from the radiative decay $J/\psi\to \gamma\eta\pi\pi$ from DM2 at Orsay, with hints at a pseudoscalar identification~\cite{Augustin:PhysRevD.42.10,Augustin:PhysRevD.46.1951}. Further evidence came in partial-wave analyses of data from other $\pi^- p$ experiments at KEK~\cite{Ando:1986bn, Fukui:1991ps} and Brookhaven~\cite{Manak:2000px, Adams:2001sk}.  Brookhaven Experiment E852 observed the \etax\ along with the \fx\ in the reaction $\pi^- p \to \eta \pi^+ \pi^- n$ at 8.45~GeV$/c$. The results were extracted from an isobar-model phase-shift analysis of the $\eta\pi^+\pi^-$ system. They found a $0^{-+}$ resonance (the \etax) to have a width of around 70~MeV$/c^2$ and a mass of about 1275~MeV$/c^2$, with an integrated production cross section two to three times that of the \fx. Concurrent to its discovery were predictions by Cohen and Lipkin~\cite{Cohen197916} that the first radial excitations of the $\eta$ and $\etapr$ mesons should lie in the 1200-1500 MeV/$c^2$ mass region.  Enumerating these states is relevant to the search for non-$\bar{q}q$ mesons in this mass range.   

With two fairly narrow mesons occupying the same mass range, it is interesting to determine which state is most strongly excited in exclusive photoproduction on the proton.  That is what this paper addresses. 


Three separate groups have predicted the photoproduction cross section for exclusive \fx\ production on the proton in the near-threshold energy regime within effective Lagrangian models~\cite{Kochelev:PhysRevC.80.025201, PhysRevD.80.115018, XiePrivComm}.   They will be introduced and compared to the experimental  results in Sec.~\ref{subsec:crosssections}.

Interest in the \fx\ state has  expanded beyond traditional meson physics. It has been shown that the  use of a leading-order chiral Lagrangian combined with a unitarization scheme can lead to the so-called dynamic generation of many well-known states.  In the meson sector, the scattering of Goldstone bosons off vector mesons can lead to a description of  many of the axial-vector meson resonances, including the \fx\ ~\cite{Lutz:2003fm, Roca:2005nm, Aceti:2015pma}.  In this framework, the meson is described not as a $\bar{q}q$ quark model object, but via dynamical state generation; the \fx\,  was found to have a dominant $K^*\bar{K} + c.c.$ quasi-bound molecular structure.  There has also been recent investigation of the \fx\ state on the lattice by Dudek~{\it et al.}~\cite{Dudek:2013yja} and by Geng~{\it et al.}~\cite{Geng:2015yta}.  The present results may help these studies, for instance through better determination of the width of the state. 

In this paper we present the first  photoproduction measurements of the \fx\ and/or \etax.  Overall, it will be evident that the \fx\ state is entirely dominant in this photoproduction reaction.  In Section~\ref{sec:data} we will present the experimental setup, and describe in Section~\ref{sec:yield} the meson yield determination and normalization method. Section~\ref{sec:acceptance} will present the efficiency and acceptance calculations, and Section~\ref{sec:normalization} will discuss the photon flux normalization.  Section~\ref{sec:systematic} will discuss systematic uncertainties in the results.   Our experimental results will be shown in Section~\ref{sec:results}, discussing precise mass and width of the state, the differential cross sections, and the branching ratios.  A spin-parity determination from amplitude analysis of the Dalitz distribution is made.  Finally, we compare our results to the world data and theoretical predictions available for the \fx. Conclusions are summarized in Section~\ref{sec:conclusions}.


\section{Experimental Procedures and Data Analysis}
\label{sec:data}

The data were obtained in the summer of 2004 using the CLAS system~\cite{Mecking2003513} located in Hall~B at the Thomas Jefferson National Accelerator Facility in Newport News, Virginia during the ``g11a'' data taking period. A 4~GeV electron beam on a gold-foil radiator of 10$^{-4}$ radiation lengths produced real photons via bremsstrahlung. The photon energies were determined, or ``tagged'', by measuring the recoiling electrons with a dipole magnet and scintillator hodoscopes~\cite{Sober2000263}. The photon energy range was 20\%-95\% of the electron beam energy and the photon energy resolution was about $~0.1\%$ of the photon energy. The target was a cylinder of liquid hydrogen, 40~cm in length and 4~cm in diameter. Target temperature and pressure were monitored throughout the experiment, such that the density was determined with an uncertainty of $\pm 0.2\%$.

The CLAS detector was segmented in six  azimuthally-symmetric sectors around the beam line.  Charged particles were tracked in each sector using three sets of drift chambers through the nonuniform toroidal magnetic field~\cite{Mestayer200081}. Charged particles with laboratory polar angles from 8$^\circ$ to 140$^\circ$ could be tracked over approximately 83\% of the azimuthal range.  Surrounding the target cell were 24 scintillator paddles comprising the start counter used in the event trigger~\cite{Sharabian2006246}.  A set of 342 scintillators, 57 per sector, located outside the magnetic field region was used in the event trigger and during offline analysis to determine the time of flight (TOF) of charged particles~\cite{SC_nim}. The momentum resolution of the detector was between 0.5\% and 1.5\%.  Other CLAS components, such as the \v{C}erenkov counters and the electromagnetic calorimeters, were not used in this analysis.  

Events were collected by requiring two charged tracks in different sectors of CLAS plus a coincident signal from the photon tagger.  The data acquisition rate for physics events was about 5 kHz, resulting in \about20 billion events or 21 TB of data. The rather ``open'' trigger in this run accumulated data simultaneously for many different photoproduction reactions, allowing for commonalities and cross-checks in the subsequent analysis, including timing and pulse-height calibrations, particle identification, and flux normalization.  A set of calibrated events containing a minimum of two positively charged tracks and at least one negatively charged track was selected.


The selection of $\gamma p \to \fx p$ events started from this reduced data set. Events consistent with the reactions
 \begin{equation}
 \gamma p \rightarrow \left\{
   \begin{array}{l}
           p \pi^+ \pi^- (\eta),\\
           p \pi^+ \pi^- (\gamma),\\ 
           p \pi^+ K^-  (K^0), \\
           p \pi^- K^+ (\bar{K}^0)
    \end{array}
    \right.
\end{equation}
were identified using kinematic fitting and time-of-flight selections.  The particles in parentheses were missing
in one-constraint (1C) kinematic fits to the reactions; confidence-level cuts were placed at 10\%.

Some events could pass the kinematic fit with one or more tracks assigned the wrong mass identity. To reject such incorrect assignments a cut was made on
\begin{equation}
  \Delta TOF = TOF_{calc} - TOF_{meas},
\end{equation}
which is the difference between the calculated time of flight for a given particle hypothesis, $TOF_{calc}$, and the measured time of flight, $TOF_{meas}$, for the track. $TOF_{calc}$ was measured for a given particle hypothesis according to
\begin{equation}
  TOF_{calc} = \frac{L}{c} \sqrt{1 + \left(\frac{mc^2}{pc}\right)^2},
\end{equation}
where $L \sim 4$m was the measured path length of the particle from the target to the TOF scintillator, $c$ is the speed of light, $m$ is the mass according to the particle hypothesis and $p$ is the measured magnitude of the momentum. $TOF_{meas}$ is
\begin{equation}
TOF_{meas} =  t_{SC} - t_\gamma,
\end{equation}
where $t_{SC}$ is the time when the particle was detected in the \clas\ TOF scintillators and $t_\gamma$ is the time when the incident photon was at the reaction vertex. The difference, $\Delta TOF$, is close to zero for tracks for which the mass hypothesis is correct. For the channels without kaons, typical CLAS selections were made with  $\Delta TOF = \pm 1.0$~nsec for at least two of the three tracks, though somewhat wider for low-momentum protons for which energy-straggling was significant.  The accelerator beam structure had a 2.0~nsec micro-structure, so this timing cut effectively constrained all particle tracks to be associated with the same event.  This criterion sufficiently reduced background while allowing for cases in which timing information was poor for one track, which reduced signal losses.  For  channels with kaons the selection criterion was strengthened, since real kaons were dwarfed in number by the number of mis-assigned protons and pions before the timing cuts were made.  We required all three tracks to pass the appropriate $\Delta TOF$ cuts for the kaon-containing decay modes, $\gamma p \to p \pi^- K^+ (\bar{K}^0)$ and $\gamma p \to p \pi^+ K^- (K^0)$.

\section{Yield Extraction}
\label{sec:yield}

\begin{figure}[htbp]
  \begin{overpic}[width=0.50\textwidth,tics=10]{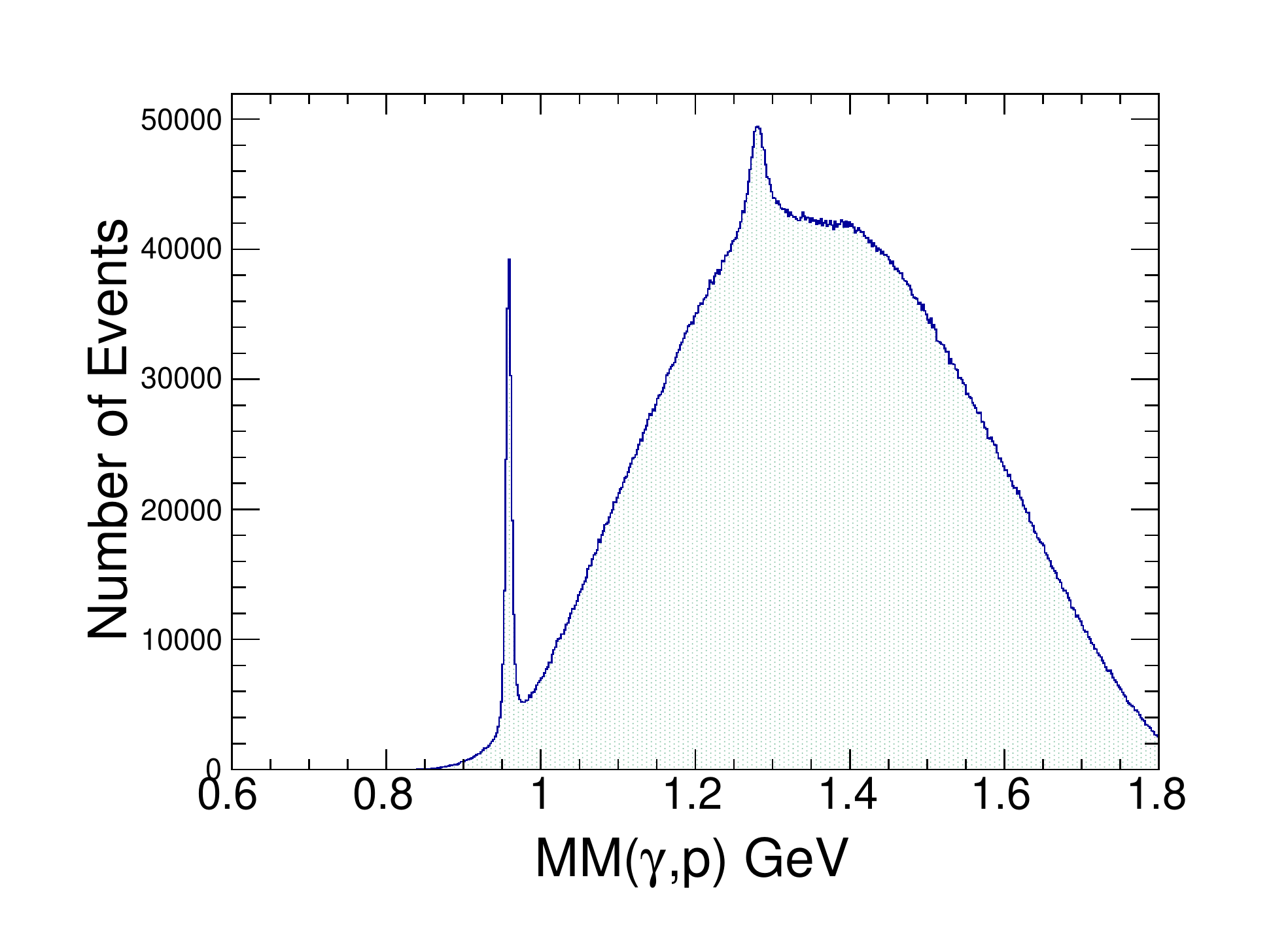}
    \put(39,57){\large \etapr}
    \put(62,59){\large \fx}
  \end{overpic}
  \caption{Missing mass off the proton  for the $\eta\pi^+\pi^- p$ final state summed over the full kinematic range. The \etapr (958) and \fx\
    mesons are visible. The \fx\ is seen atop a substantial multi-pion
    background.
  \label{fig:MMp_etapipi}}
\end{figure}

Figure~\ref{fig:MMp_etapipi} shows the missing mass $m_x$  in the reaction $\gamma p \rightarrow x p$ when the final state is kinematically fit to $\pi^+ \pi^- p (\eta)$, where the $\eta$ is the missing particle.  One clearly sees the $\eta^\prime(958)$ and the meson we will eventually conclude is the \fx. There is no signal bump visible corresponding to the $f_1(1420)$, unlike in $pp$ central production~\cite{Barberis:1998by}.  The large broad background is from events with four or more pions in the final state.  Some combinatoric background comes from $\eta$-decay pions or pions from other $f_1(1285)$ and $\eta^\prime$ decay modes.  The overall signal to background ratio for the \fx\ is approximately 1:6.  Figure~\ref{fig:MMp_etapipi} shows the background is peaked at about the same location as the  \etapipi\ decay mode of the \fx, which was a challenge when extracting yields in the lowest-statistics bins.  Particle yields for both the $\eta^\prime(958)$ and the \fx\ were determined by two methods, to be described next, in each kinematic bin of $\gamma p$ center-of-masss energy, $W$, and cosine of the meson production angle, $\cos\Theta^{c.m.}$.  The $\eta^\prime(958)$ cross sections compared to published data serve as a check of these methods.

\begin{figure}[htbp]
 \subfloat{%
    \begin{overpic}[width=0.48\textwidth,height=0.25\textheight]{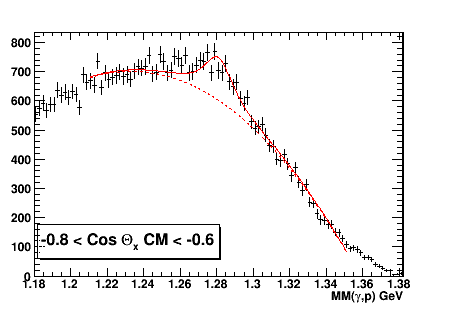}
      \put(75,55){$(a)$}
       \put(-2,37){\makebox(0,0){\rotatebox{90} {Number of Events}}}
    \end{overpic}
    \label{fig:W235_ct-0.7_voigt_mmp}
  } \\
  
  \subfloat{%
    \begin{overpic}[width=0.48\textwidth,height=0.25\textheight]{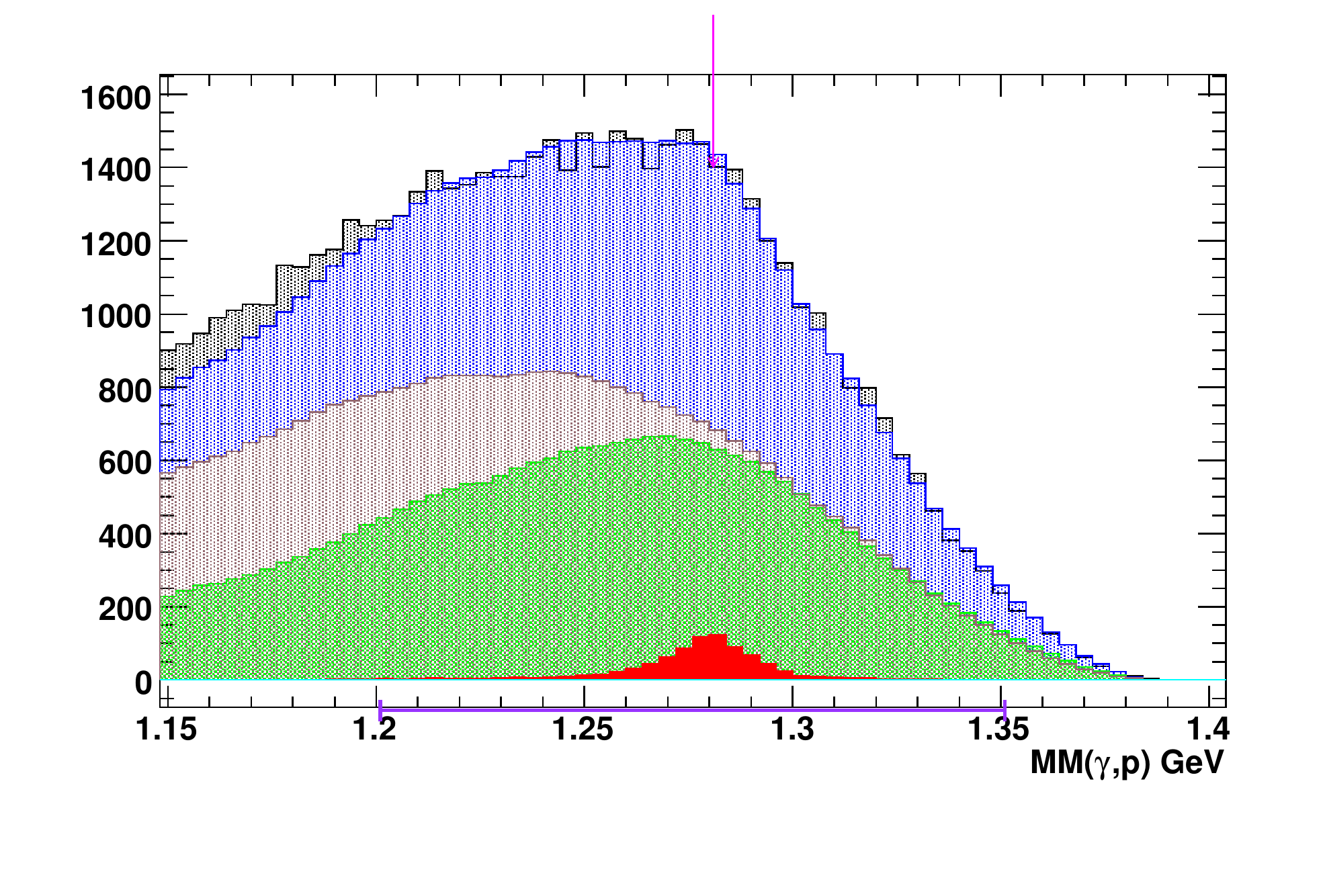}
      \put(75,55){$(b)$}
       \put(2,40){\makebox(0,0){\rotatebox{90} {Number of Events}}}
    \end{overpic}
   \label{fig:W235_ct-0.7_mcbkgd_mmp}
 }   
  \vspace{-0.5cm}
  \caption{(Color online) 
    Example fits for two methods of treating spectra of missing mass
    off the proton for $\gamma p \to p\eta \pi^+\pi^- $ in the same bin $W =
    2.35$~GeV and  $\cos \Theta^{c.m.} = -0.7$. 
    (a) Voigtian signal with polynomial
    background (solid red).  The fit region, signal yield, and smooth background (dashed red) are seen; 
    (b)  Monte Carlo template
    method with smoothed 4~MeV wide bins. The \fx\ signal is the lowest
   histogram (red), with the $p \rho\pi\pi$ background (green), the
    $f(1370) p$ background (brown), the sum of the three simulated
    spectra (blue), and the data (grey).
    }
  \label{fig:etapipi_mmp_example_fits}
\end{figure}

Both yield extraction methods are illustrated in Fig.~\ref{fig:etapipi_mmp_example_fits}.  The first method used a least-squares fit to a Voigtian signal lineshape plus a third-order polynomial background function.  The Voigtian distribution
\begin{equation}
  V(E;M,\sigma,\Gamma) = \int_{-\infty}^{\infty} G(x';\sigma) L(E-x';M,\Gamma)\, dx'
\end{equation}
is a convolution of a non-relativistic Breit-Wigner (Lorentzian) $L$ with a Gaussian $G$ whose width, $\sigma$, was set to the calculated mass resolution. Our method was to leave the width $\Gamma$ and mass $M$ of the meson free in the high-statistics bins and to fix them to the overall ``best'' values in the low-statistics bins.  We fixed $\sigma$ on a bin-by-bin basis to  values between 3 and 6~MeV obtained from the Monte Carlo simulation of CLAS discussed in Sec.~\ref{sec:acceptance}.  Figure~\ref{fig:W235_ct-0.7_voigt_mmp} shows an example fit of the \fx\ signal in one of the more challenging bins for $\gamma p \to p\pi^+\pi^- \eta$. The fit range for most bins was 140~MeV$/c^2$ in \mmp, centered at 1281~MeV$/c^2$.

The second method used Monte Carlo simulation of  the \fx\ signal and a set of simulated multi-pion reactions to approximate the background shape seen in the MM$(\gamma,p)$ spectrum in the \etapipi\ channel, including combinatorics. A category of background events not rejected by the event selection criteria contained a proton plus four pions, in which the ``extra'' pions were either neutral or undetected charged pions. Both of these types of final state passed our kinematic fit to $p \pi^+ \pi^-(\eta)$ if the invariant mass of the ``missing'' pions was near the mass of the $\eta$. The reactions $\gamma p \to p \pi\pi\pi\pi$, $\gamma p \to p \rho\pi\pi$, $\gamma p \to \Delta\pi\pi\pi$ and $\gamma p \to p f_{0}(1370)$, all of which have a final state of a proton and four pions, were generated according to phase space and passed though the analysis. The inclusion of the $\Delta$ and $\rho$ modes was motivated by the evident presence of these backgrounds in the invariant mass spectra $IM(p \pi)$ and $IM(\pi \pi)$. While these reactions did not represent all possible physics backgrounds, these four pion final states were chosen to populate the kinematic space of the data.

The phase space for these background reactions was much larger than for the signal reaction, resulting in low statistics in the signal region, especially in the highest $W$ bins. To compensate for this we implemented a smoothing algorithm using 4 MeV-wide bins in missing mass and using a quadratic polynomial smoothed over a 100~MeV range in $MM(\gamma,p)$. This was done iteratively to obtain the smooth background shapes shown in the Fig.~\ref{fig:W235_ct-0.7_mcbkgd_mmp}.
Fits were made in each kinematic bin were fit with the set of smoothed backgrounds and the signal Monte Carlo spectra.  The meson yield from this method was simply the integrated signal Monte Carlo scaled by its fit coefficient. Figure~\ref{fig:W235_ct-0.7_mcbkgd_mmp} shows a fit to the \fx\ signal in one bin for $\gamma p \to p\pi^+\pi^- \eta$.  The data shown in both panels is for one of the statistically quite marginal bins in which the two methods nevertheless produced  consistent results.

\begin{figure*}[]
  \centering
  \makebox[\textwidth][c]{
    \subfloat{%
        \begin{overpic}[angle=90,width=0.48\textwidth,height=0.25\textheight]{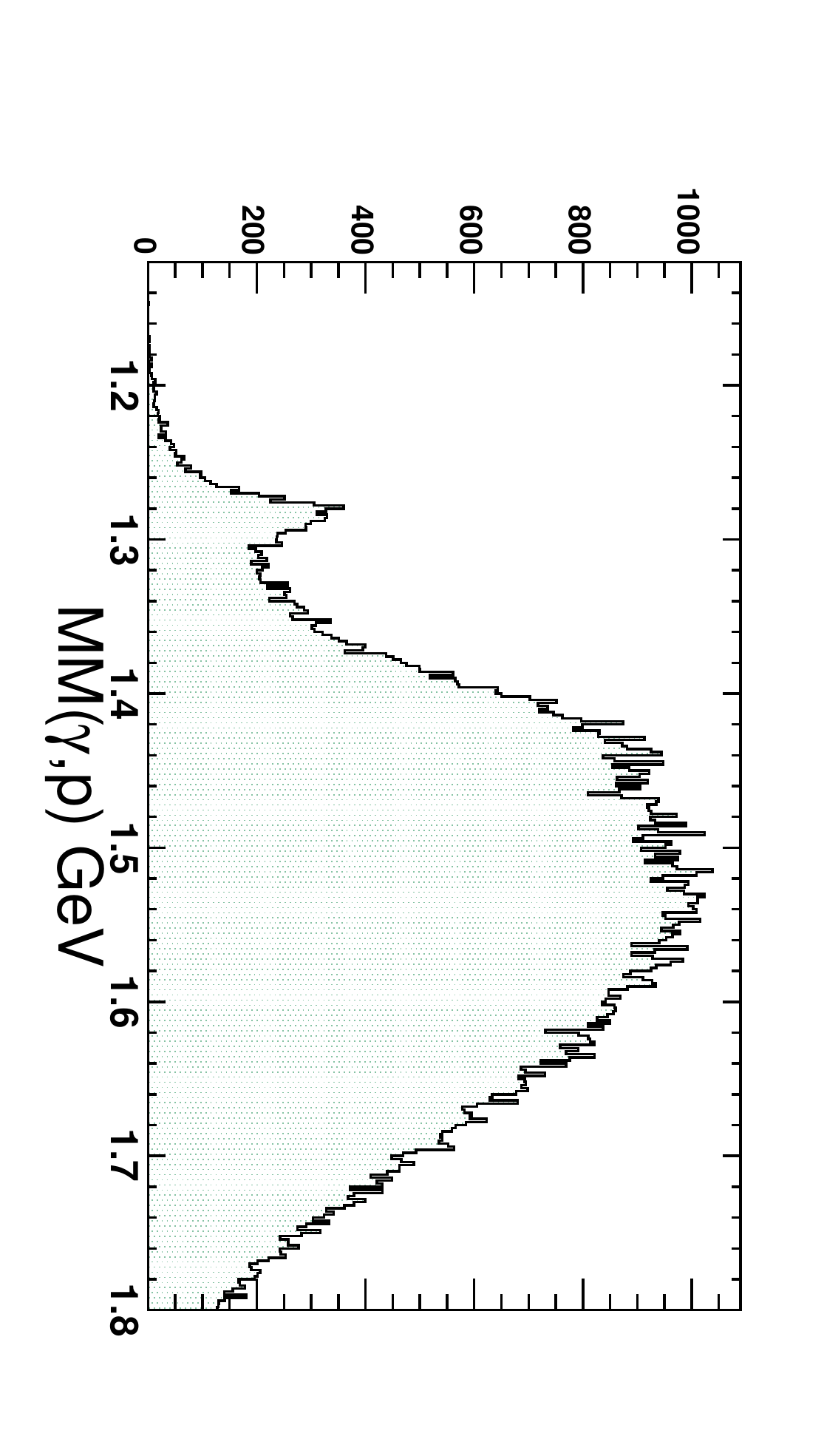}
       \put(25,50){$(a)$}
       \put(8,35){\makebox(0,0){\rotatebox{90} {\large{Number of Events}}}}
       \end{overpic}
      \label{fig:mmp_x1280_kkpi-a}%
    }
    \subfloat{%
      \begin{overpic}[angle=90,width=0.48\textwidth,height=0.25\textheight]{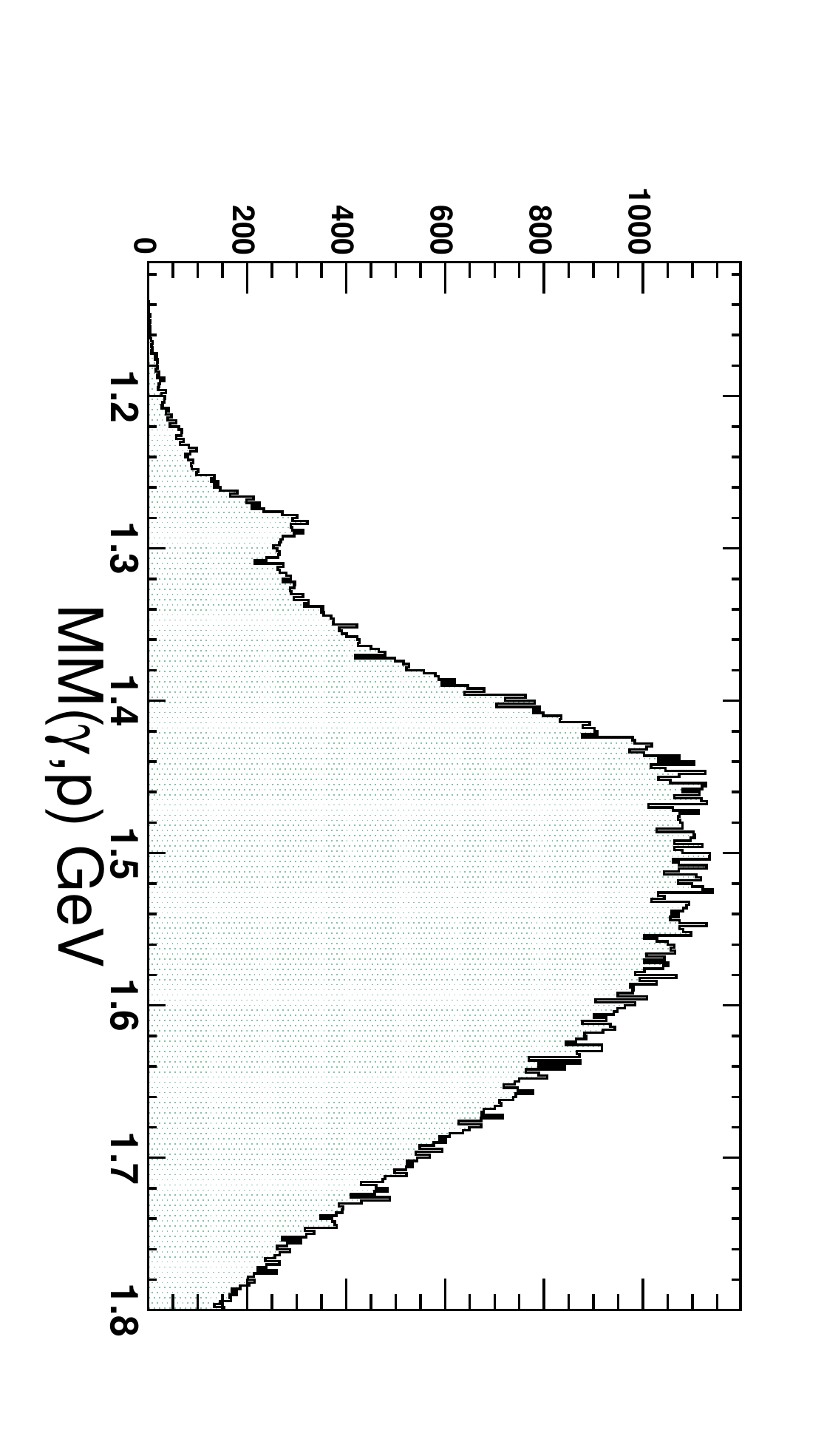}
      \put(25,50){$(b)$}
       \put(8,35){\makebox(0,0){\rotatebox{90} {\large{Number of Events}}}}
      \end{overpic}
      \label{fig:mmp_x1280_kkpi-b}%
    }
  }
  \caption[]
 {Missing mass off the proton for \subref{fig:mmp_x1280_kkpi-a} $\gamma p \to p  \pi^- K^+ (\bar{K}^0)$ and \subref{fig:mmp_x1280_kkpi-b} $\gamma p \to p \pi^+ K^-  
(K^0)$.  The small bumps at 1.28 GeV show the meson of interest. These spectra are summed over the full kinematic  domain. }
  \label{fig:mmp_x1280_kkpi}
\end{figure*}

\begin{figure}[htb]
  \centering
  \begin{overpic}[width=0.50\textwidth]{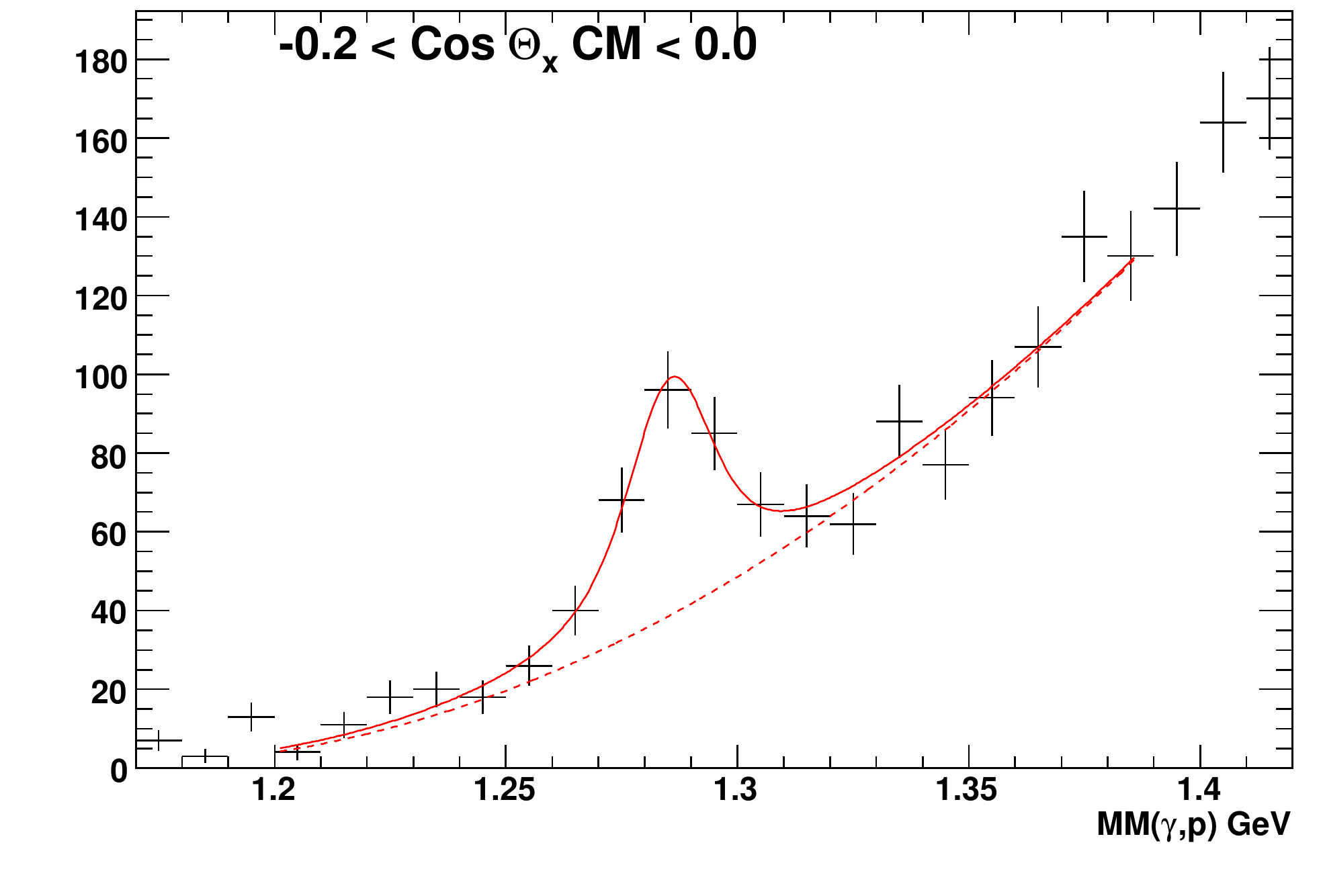}
       \put(2,40){\makebox(0,0){\rotatebox{90} {\large{Number of Events}}}}
  \end{overpic}
  \caption{Combined missing mass off the proton spectra for $\gamma p  \to p  \pi^- K^+(\bar{K}^0)$ plus $\gamma p \to p  \pi^+ K^-(K^0)$.  Example Voigtian-plus-polynomial fit (total is solid red, polynomial is dashed) of \fx\ yield in \kkpi\ at $W = 2.45$~GeV and  $-0.2<\cos\Theta^{c.m.}<0.0$. 
     }
  \label{fig:mmp_x1280_kkpi_sample_fit}
\end{figure}

For the \kkpi\  decay modes of the \fx\ the background was smaller and less-rapidly changing so we only used the Voigtian yield extraction method.  This is illustrated in Fig.~\ref{fig:mmp_x1280_kkpi}.  The total statistics for the kaon channels was smaller than for the $\eta\pi\pi$ decay mode.  This was handled by combining these two charged kaon modes prior to yield extraction.  Fits were then made to the summed data in each kinematic bin; an example is shown in Fig.~\ref{fig:mmp_x1280_kkpi_sample_fit}. The fits used a Voigtian line shape plus polynomial function, with the mass  and width fixed to the best values obtained from the spectrum integrated over all production angles at a given $W$.

\begin{figure}
  \centering
  \begin{overpic}[width=0.50\textwidth]{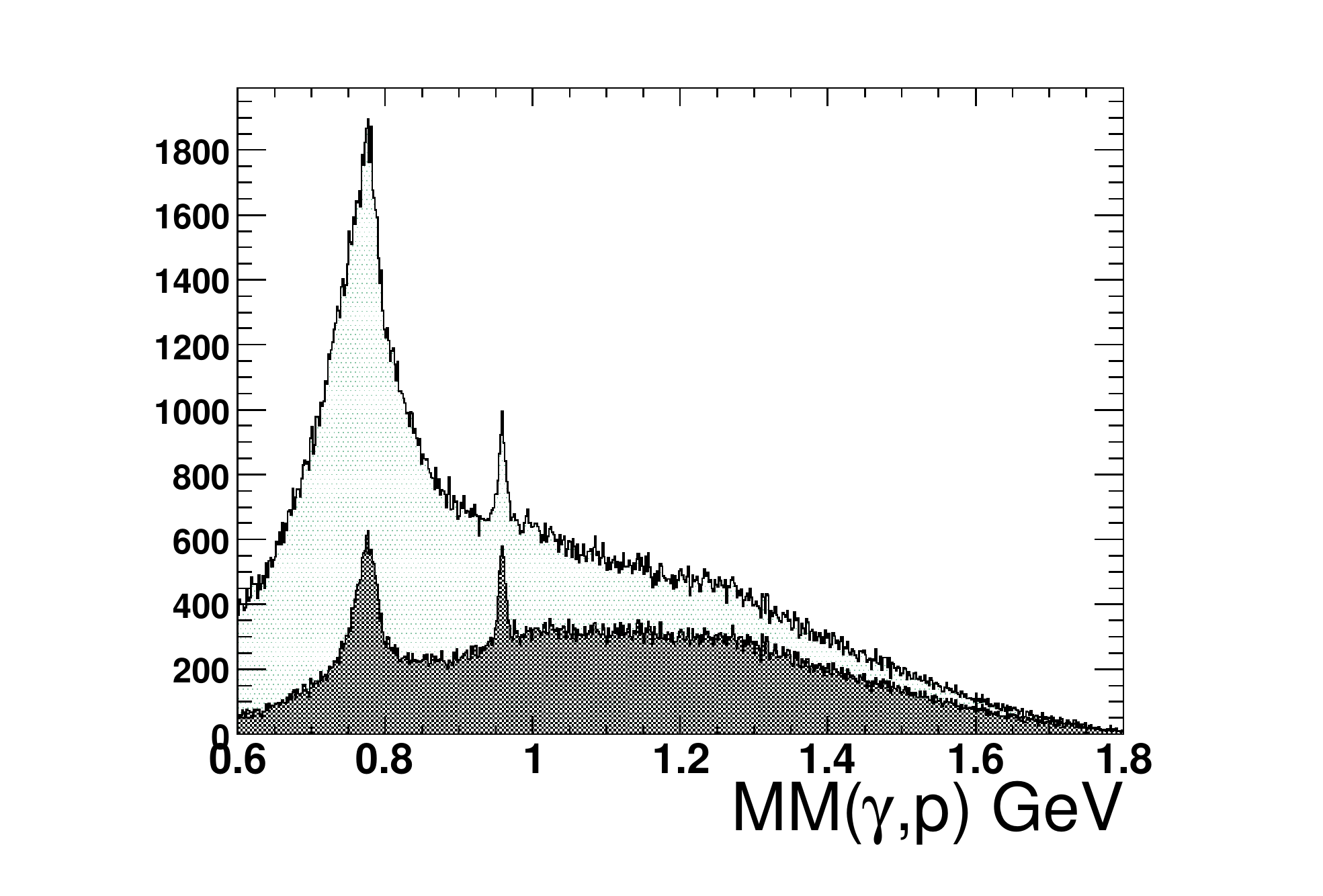}
         \put(8,35){\makebox(0,0){\rotatebox{90} {\large{Number of Events}}}}
  \end{overpic}
  \caption {\MMp\ for $\gamma p \to p \pi^{+} \pi^{-} (\gamma)$ events after kinematic fitting and confidence level cut for one run (open histogram). A strong
    signal for $\omega / \rho^0$ near 800\,\mmunit\ remains, as well as the \etapr(958) and a slight hint of the broad $a_2(1320)$. The
    filled histogram shows the effect of  a minimum missing transverse momentum in reducing background. No significant \fx\ signal is seen at this stage. }
  \label{fig:MMp_rhogamma}
\end{figure}

The last decay mode to extract was the channel $\fx \to \gamma\rho^0$, where the $\rho^0$ decays $\sim100\%$  to $\pi^+\pi^-$. As the $\rho^0$ is quite wide at $\Gamma \approx 150$~MeV we did not impose any cuts on the invariant mass of the two-pion system.   The kinematic fit to $\gamma p \rightarrow p \pi^{+} \pi^{-} (\gamma)$ selected events with zero missing mass and any missing momentum.  The confidence-level cut  alone did not distinguish between events with no missing particle and  signal events with a photon. The $\eta^\prime(958)$, \fx, and \etax\ mesons do not decay to $\pi^+\pi^-$ alone due to parity, so it was desirable to remove such events to improve our signal to background ratio.  To separate signal events with a missing photon from exclusive $\gamma p \rightarrow p \pi^{+} \pi^{-}$  events we imposed a minimum missing momentum cut. The kinematic fit had more freedom to adjust momentum along the beam direction than perpendicular to it. This was due to the uncertainty in the incident photon energy, along with the possibility of having chosen the wrong in-time photon from the multiplicity of photons in a given event.  A $p \pi^{+} \pi^{-}$ event with no missing particle cannot have any appreciable transverse momentum $P_{\perp}$. To determine an effective minimum transverse momentum cut, we examined the spectrum for $\etapr$ events as it has a smaller breakup momentum and sufficient statistics to fit the signal in the low-$P_{\perp}$ region. From this study, we required events to have $>40$ MeV/c missing transverse momentum to select events with a missing photon.  Figure~\ref{fig:MMp_rhogamma} shows the effect of this selection.

After removing events with small missing transverse momentum, there was still a sizable $\omega$ signal seen in the \mmp. The dominant decay mode of the $\omega$ is into $\pi^+\pi^-\pi^0$. The remaining peak suggested that events with a missing $\pi^0$ were being pulled into the  missing-$\gamma$ kinematic fit.  To remove these events from the $\gamma p \rightarrow p \rho^{0}\gamma$ sample we performed a second kinematic fit to the missing $\pi^{0}$ hypothesis. Events which passed this fit with a confidence level of more than $0.01$ were removed from the $\rho^{0}\gamma$ data sample.

\begin{figure}[htb]
 \centering \subfloat{
      \label{MMp_rhogamma_final-a}
    \begin{overpic}[width=0.50\textwidth,tics=10] {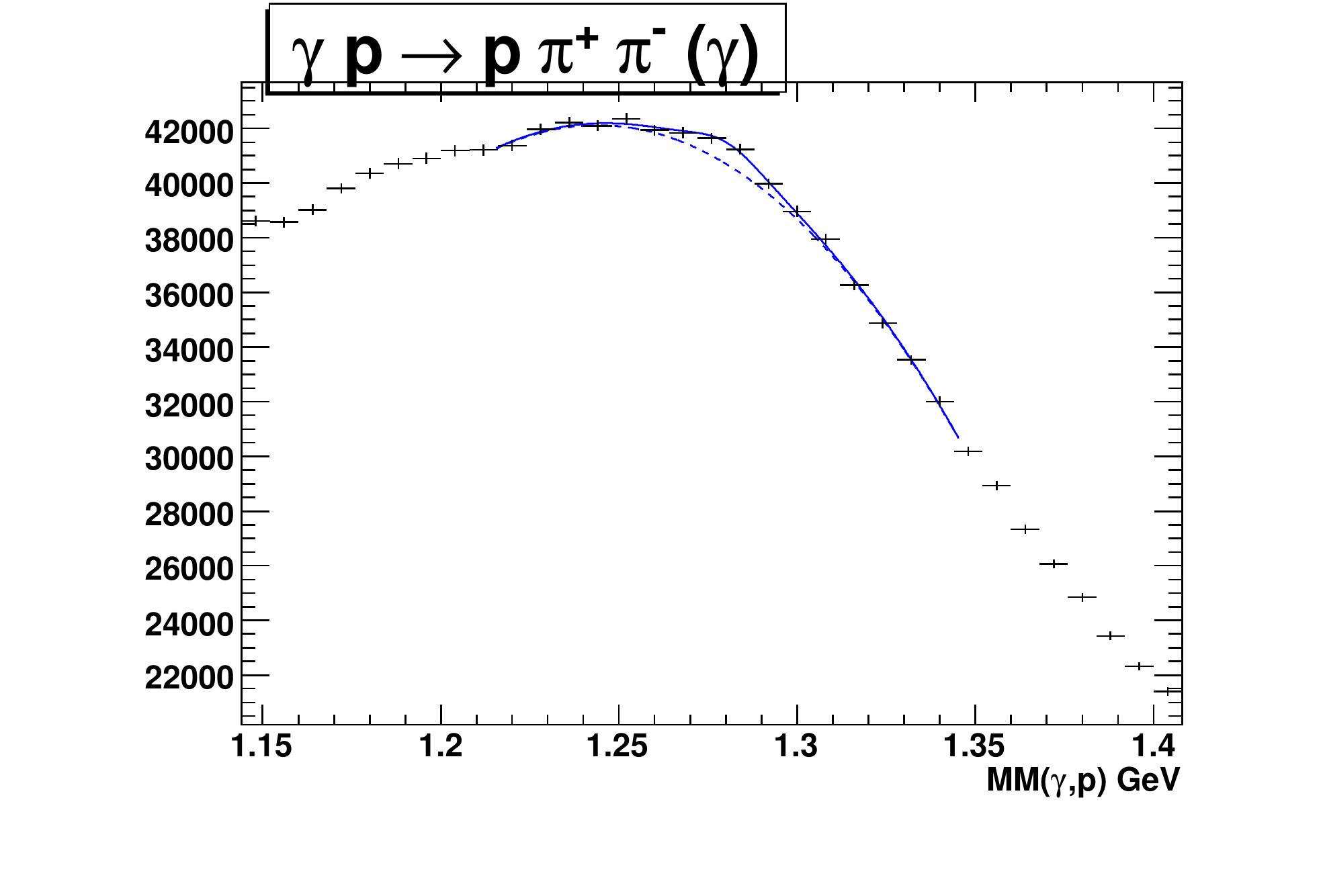}
     \put(75,50){(a)}
     \put(5,35){\makebox(0,0){\rotatebox{90} {\large{Number of Events}}}}
   \end{overpic}
      }
   \vspace{-20pt}
    \subfloat{
      \label{MMp_rhogamma_final-b}
   \begin{overpic}[width=0.5\textwidth,tics=10] {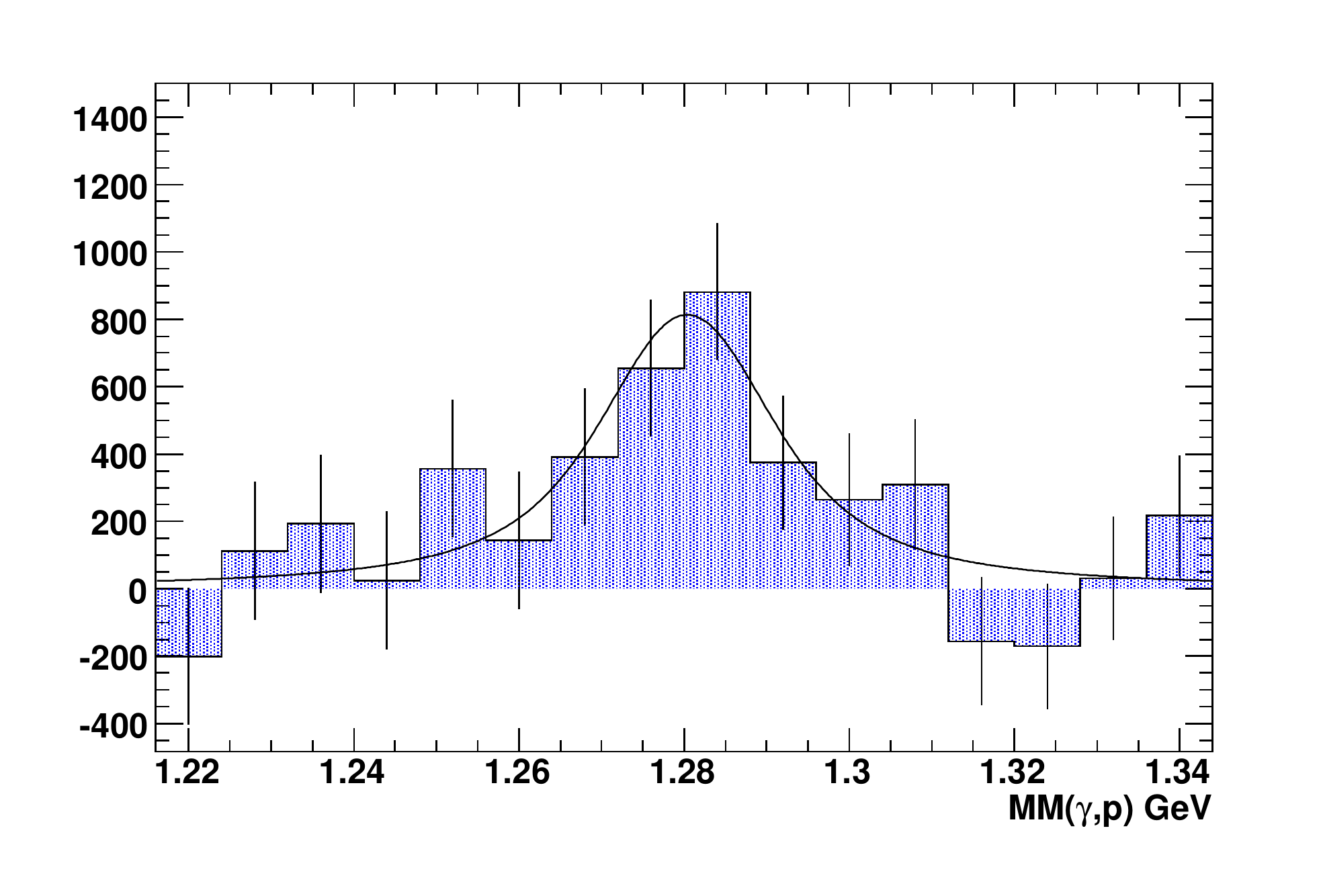}
     \put(75,50){(b)}
     \put(0,35){\makebox(0,0){\rotatebox{90} {\large{Number of Events}}}}
   \end{overpic} 
    }
    \vspace{-0.5cm}
    \caption[] {\subref{MMp_rhogamma_final-a}~\MMp\ for $\gamma p \to
      p \pi^+ \pi^- (\gamma)$ fitted with a Voigtian signal shape plus a polynomial
      background.  \subref{MMp_rhogamma_final-b}~Spectrum after
      background subtraction using the polynomial parameters from the
      fit. }
  \label{fig:MMp_rhogamma_final}
\end{figure}

After these steps to reduce background from other final states from $\gamma p \to p \pi^+ \pi^- (\gamma)$ events, we found the spectrum to have a small but discernible signal at 1280~\mmunit, as shown in Fig.~\ref{fig:MMp_rhogamma_final}.  This spectrum represents the totality of our data set for this decay mode.  The statistics and the signal-to-noise ratio were too poor to bin the data for extraction of differential cross sections. We instead fitted the total missing mass spectrum to determine the total yield of $\fx$ in the \rhogamma\ final state for calculating branching ratios.

\section{Acceptance, Efficiency, \& Normalization}
\label{sec:acceptance}
The proportion of events lost due to detector inefficiencies, geometric acceptance and analysis cuts was computed as a function of energy and angle using a well-tested computer simulation of \clas\ called GSIM~\cite{holtrop_gsim}. Simulated events were processed through the same event reconstruction and analysis software as used for the actual data. We removed from our analysis detector regions where the simulation did not accurately reproduce the data as described earlier.  Through studying the simulated events we obtained acceptance factors as a function of the kinematic binning of the differential cross sections, as well as a measure of the expected experimental resolution in mass spectra.

\subsection{Monte Carlo Event Generation}
\label{sec:Signal_MC}

20 million events were generated each for $\eta^\prime\rightarrow\eta\pi\pi$ and     $\eta^\prime\rightarrow\rhogamma$ decays.  For the \fx, 10 million events were generated per decay mode.  The event generator included the bremsstrahlung energy distribution and a $t$-slope in the meson production angle, as is known to apply for the \etapr\ and estimated for the \fx.  A small flat baseline flux was included to ensure an adequate number of events were produced in the backward angle bins at high energy, where the combination of the $t$-slope and bremsstrahlung distributions otherwise led to a very small number of events generated.  Decays leading to combinatoric background were included.  Decays were calculated according to 3-body phase space, excepting the \rhogamma\ final state, where first the two-body decay into $\rho^0$ and $\gamma$ was generated followed by the decay of the $\rho^0$ into $\pi^+ \pi^-$.  The line shape of the $\rho^0$ included the easily-seen $\sim 25$~MeV reduction in the centroid of the $d\sigma / dm_{\pi\pi}$ distribution  due to the Drell mechanism~\cite{Soding:1965nh}. The apparent mass of the $\rho$ influences the momentum distribution of the missing photon, the transverse component of which was used for a background reduction cut, as mentioned earlier.

The simulation did not model the CLAS hardware trigger.  Inefficiencies arose if a track did not meet the trigger discriminator threshold for the TOF PMTs or if the timing windows between detector components did not match.  A study was performed using $\gamma p \to p \pi^+ \pi^-$ exclusive events~\cite{bellis_clas_note:2006-017}, in which any of the three tracks could be predicted from the other two in order to see whether the detector found the third track.  The resultant map of trigger efficiency as a function of charge, TOF paddle, and track momentum was used in simulation of the trigger for this data set, which required two or more tracks in different sectors to trigger an event.

The fully exclusive reaction $\gamma p \rightarrow \pi^+ \pi^- p$ was used to map both single-track momentum corrections and detector inefficiencies in detail~\cite{bellis_clas_note:2006-017}.  Fiducial cuts were applied on the momenta and angles of the tracks to select events from the well-understood regions of the detector. An algorithm smeared the track angle and momentum of the simulated events in accordance to kinematic fit results in exclusive  $\gamma p \to p \pi^+ \pi^-$ events, as detailed in Ref.~\cite{Williams:th}.  The overall agreement was excellent between the experimental apparatus and simulation. 

\subsection{Acceptance Calculation}

Simulated events were reconstructed with the same analysis code used  for real  events. The acceptance of \etapr\ and \fx\ events, including both the detector efficiency and the signal loss from the event selection criteria, was then computed.  For each final state the acceptance for a given energy and production angle is
\begin{equation}
  \mathcal{A}_{\mathrm{CLAS}}(W,\cos\Theta^{c.m.}) = \frac{\mathcal{N}_{acc}}{\mathcal{N}_{gen}},
\end{equation}
where $\mathcal{N}_{acc}$ is the number of accepted events and $\mathcal{N}_{gen}$ is the number of generated events in that kinematic bin.

The acceptance for the physics ``signal'' processes did not include simultaneous calculation of the four-pion backgrounds.  The calculation of $\mathcal{N}_{acc}$ was performed using the methods of yield extraction for each meson and decay modes described in Sec.~\ref{sec:yield}.  The acceptance of the \etapipi\ decay mode increases with $W$, especially at mid to backward angles. For the \rhogamma\ mode the situation was reversed: it has the highest acceptance at the lowest energy bin and decreases with increasing $W$.  For the \fx\  acceptance, we performed a Voigtian fit with the width $\Gamma$ fixed to 18~MeV, which was the experimentally observed value and therefore also the input value used in the Monte Carlo event generator. The resulting $\sigma$'s from these Voigtian fits were taken to be the experimental resolution and used as input for the yield extraction fits to the data discussed in Sec.~\ref{sec:yield}. The trends in acceptance for $\fx \to \etapipi$ are similar to those seen for $\etapr \to \etapipi$, increasing with energy and with the highest values at central polar angles. The maximum acceptance value was about 10\%.  For the \kkpi decay modes, the acceptance is smaller with a maximum of about 4\% in the highest $W$ bins.  The Monte Carlo event generation was iterated to better match the observed differential cross section for the \fx, as discussed later; this was also necessary to quantify the resulting systematic uncertainty on the acceptance, particularly for the $\gamma\rho^0$ decay mode.

\section{Normalization}
\label{sec:normalization}
CLAS photoproduction measurements are normalized using a calculation of the number of electrons that hit the hodoscope of energy-defining scintillators in the photon tagger.  These detectors are part of the event trigger, but their asynchronous hit rate is closely related to the number of photons tagged at a given energy.  Corrections based on measurements are made for losses between the photon tagger and the physics (hydrogen) target.  The same photon flux calculations used for the present measurement have been used for several previously-published results from this data set~\cite{Williams:2009ab,Williams:2009yj,McCracken:2009ra,Dey:2010hh,Moriya:2013eb,Moriya:2013hwg,Dey:2014tfa}.  

\section{Systematic uncertainties on cross sections}
\label{sec:systematic}
To estimate systematic uncertainty on the event yields, we varied the fitting conditions and compared the results of the two methods discussed in Sec.~\ref{sec:yield}.  Five variations of the fits using  the Voigtian method for $\fx \to \etapipi$ were tested. The range of \mmp\ was varied by $-10\;{\rm MeV}$, $+10\;{\rm MeV}$ and $+20\;{\rm  MeV}$. The background polynomial was increased from third to fourth and fifth order.  These variations gave a bin-dependent uncertainty from 1.9~to~4.1\% of the yield of \fx\ in \etapipi.

For the $\fx \to \kkpi$ case, we again varied the fit range and polynomial background from the central conditions in the same manner as above. This gave an average uncertainty from 2.2~to~5.7\% of the yield to the  \kkpi final state.

For the $\fx \to \rhogamma$ decay the experimental signal-to-background ratio was too small to allow binning in center-of-mass energy and production angle, so only a total yield was extracted. The width and mass of the \fx\ were fixed to the best values found from the \etapipi\ fits, while the Gaussian width $\sigma$ in the Voigtian function was fixed  from analysis of \fx\ Monte Carlo events. To test the stability of the small \rhogamma\ signal to analysis variations, the fit range and the order of the polynomial background used in the fit were changed. The standard deviation of the yield values was found to be 22.4\% of the $\fx \to \rhogamma$ yield.

The second yield-extraction method using the simulated background gave lower yields with larger uncertainties for almost all energy and angle bins. The differential cross sections calculated by this method agreed with the values obtained via the Voigtian yield fits within the respective error bars.  Therefore, when we combined the results from the two yield extraction methods, we ascribed half the difference between them as the systematic uncertainty estimate included in the final bin-to-bin uncertainties for the differential cross sections for $\fx \to \etapipi$.  Thus, we believe to have fairly estimated the uncertainty in measuring the small \fx\ signal on top of a large background.

Systematic cross section normalization uncertainty was studied by several prior analyses of $\omega$~\cite{Williams:2009ab} and \etapr~\cite{Williams:2009yj} photoproduction from the same data set.  To estimate the systematic uncertainty due to the calculation of the photon flux, we studied the variation of \etapr\ yield in the \etapipi\ channel normalized by the photon flux for each production run in the data set.  Results were fully consistent.  Comparison of cross sections in $p\omega$, $KY$, and $p\eta$ final states extracted from ``g11a'' to previous world data including the earlier CLAS ``g1c'' data set~\cite{Bradford:2005pt} led the authors in Refs.~\cite{Williams:2009ab}~and~\cite{Williams:2009yj} to assign a global value for the flux normalization uncertainty. These comparisons assigned a systematic uncertainty of 7.3\% on the photon flux normalization and we adopt that value for this analysis.

Combining in quadrature the photon normalization uncertainty of $7.3\%$ with an uncertainty of 0.2~to~0.5\% due to photon beam line attenuation~\cite{Schumacher_clas_note:2001-010} and a $3\%$ uncertainty in the data-acquisition live-time correction~\cite{Williams:2009yj}, gives an overall systematic uncertainty for normalization of 7.9\%.

The systematic acceptance uncertainty was estimated previously~\cite{bellis_clas_note:2006-017} by an extensive empirical study of the reaction $\gamma p \to p \pi^+\pi^-$.   To estimate additional systematic uncertainty in the present reactions we used the symmetry of the six CLAS sectors. The reactions studied in this analysis are azimuthally symmetric, allowing calculation of differential cross-sections independently for each of six CLAS sectors.  To use this method, we measured the high-statistics cross sections of the \etapr\ and applied the result to the lower-statistics \fx.  The outcome was an energy-averaged 9\% systematic uncertainty due to limits in the precision of the acceptance corrections.  The same acceptance uncertainties were used in the \kkpi\ channel where there is no high-statistics reference channel analogous to the \etapr\ for the \fx\ channel.

\begin{figure*}[htpb]
  \centering
  \includegraphics[width=1.0\textwidth]{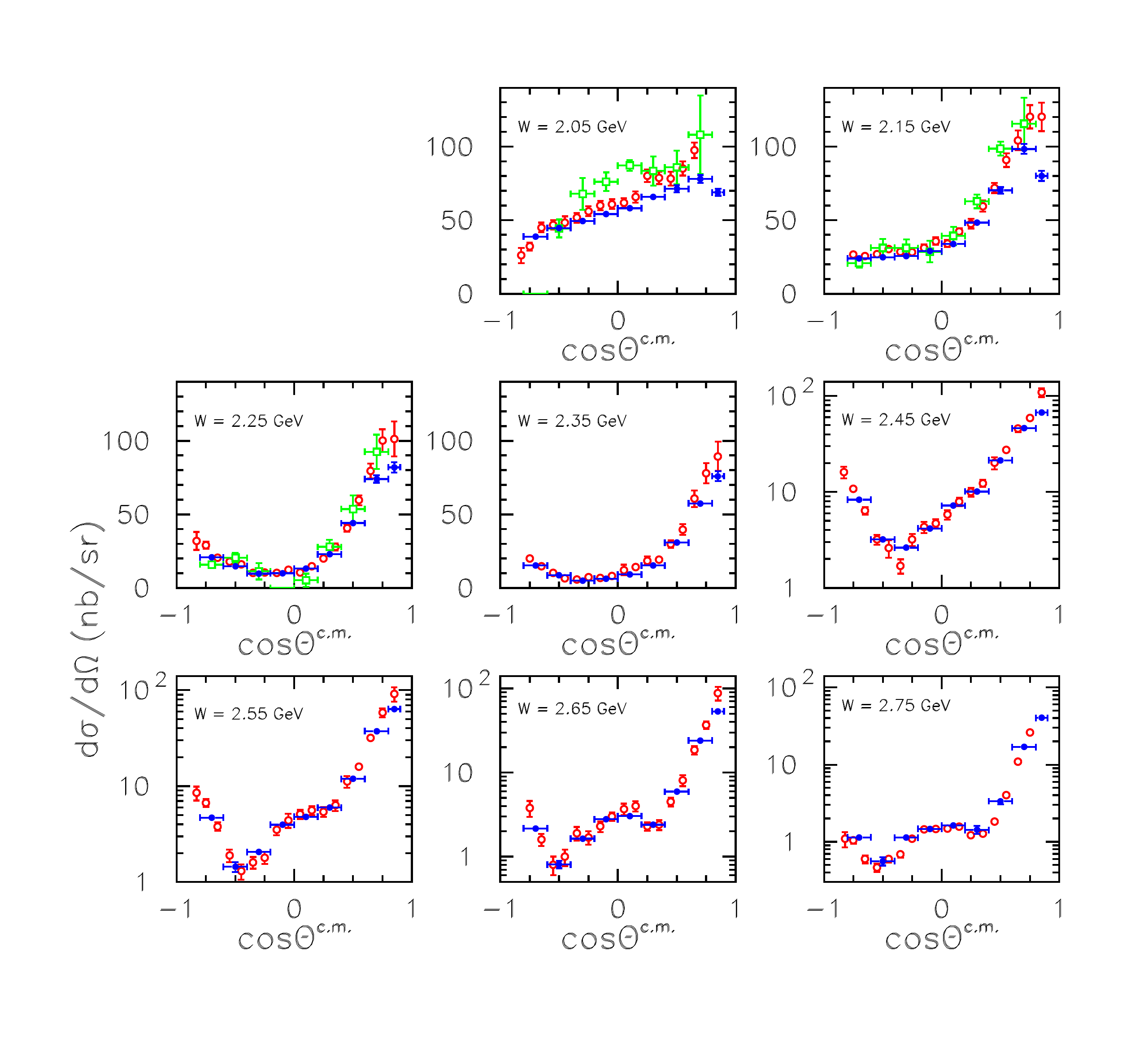}
  \vspace{-50pt}
  \caption{Differential cross sections $d\sigma/d\Omega$ for $\gamma p  \to \etapr p$ from the present measurement (solid blue points) compared to recent CLAS  published results from Ref.~\cite{Williams:2009yj} (open red circles) and Ref.~\cite{Dugger:2005my} (open green triangles).  
Note bins upwards from  $W = 2.45$~GeV are on  semi-logarithmic scales. The agreement for $W=2.35$~GeV and above, bins used later in this analysis, is very good.}
  \label{fig:R_dcs_etaprime_compare}
\end{figure*}

\begin{table}[htpb]
  \centering
  \begin{tabular}{ cccc }
    \hline\hline
    Source & Energy Bin & \multicolumn{2}{c}{Fractional Uncertainty} \\
           & $W$ (GeV)  &                                            \\
    \hline
    &      & \etapipi & \kkpi \\
    \hline
    \multirow{5}{*}{Yields} 
    &    2.35	& 0.041 & 0.032\\
    &    2.45	& 0.022 & 0.030\\
    &    2.55	& 0.026 & 0.022\\
    &    2.65	& 0.018 & 0.022\\
    &    2.75	& 0.040 & 0.057\\
    \hline
    \multirow{5}{*}{Acceptance} 
    &   2.35 &  \multicolumn{2}{c}{0.11} \\
    &   2.45 &  \multicolumn{2}{c}{0.08}\\
    &   2.55 &  \multicolumn{2}{c}{0.08}\\
    &   2.65 &  \multicolumn{2}{c}{0.11}\\
    &   2.75 &  \multicolumn{2}{c}{0.085}\\
    \hline
    Event Selection & All & \multicolumn{2}{c}{0.002}\\
    \hline
    Normalization   & All & \multicolumn{2}{c}{0.079} \\
    \hline
    \multirow{5}{*}{Combined Total}
    & 2.35& 0.14 & 0.14 \\
    & 2.45& 0.11 & 0.12 \\
    & 2.55& 0.12 & 0.12 \\ 
    & 2.65& 0.14 & 0.14 \\
    & 2.75& 0.12 & 0.13 \\
    \hline
    \hline
  \end{tabular}
  \caption {Systematic uncertainty summary for the \fx\  $d\sigma / d\Omega$ differential cross section measurements.
  }
  \label{tab:final_systematics}
\end{table}

The  acceptance value used to correct the yield of $\fx \to \rhogamma$ events was not binned in energy and angle. Due to the integration of events over a very wide kinematic space, this value is quite sensitive to any discrepancy between the physics of the reaction and the distribution of the generated Monte Carlo events. As  discussed earlier, the \fx\ Monte Carlo event generator simulated the bremsstrahlung photon energy distribution and an estimated $t$-slope. A revised sample of Monte Carlo events was  distributed according to the measured differential cross section from the first iteration to test for systematic shift in the $\fx \to \rhogamma$ acceptance. The initial Monte Carlo acceptance of $\fx \to \rhogamma$ was 2.98\% and the value for the revised empirical version is 2.48\%. We adopted the new value for our branching ratio result and used the difference between the iterations as the estimate of the systematic uncertainty on the acceptance for this decay mode.

One additional method of estimating the systematic uncertainty on cross sections was to compare results from the different decay modes.  In particular, the cross section $d\sigma / d\Omega$ for $\etapr$ calculated from the \rhogamma\ channel was $5\%$ higher, on average, than the results from the  \etapipi\ channel. This difference was used in estimating the systematic uncertainty in our calculations for the \fx\ branching ratio $\Gamma(\rhogamma) / \Gamma(\etapipi)$.

Table~\ref{tab:final_systematics} itemizes the sources of systematic uncertainty for the differential cross section results. These are the global uncertainties that apply to each of the specified bins in $W$, and over the full range of production angles.

Preliminary to presenting results for the \fx\ cross sections, we  show that the analysis methods successfully reproduce previous results.  Shown in Fig.~\ref{fig:R_dcs_etaprime_compare} are differential cross sections for the $\gamma p \rightarrow \eta^\prime p$ reaction. In eight bins in $W$ we compare the results of this analysis (blue solid points with horizontal and vertical error bars) with the CLAS-published results of Ref.~\cite{Williams:2009yj} (red points with only vertical error bars) and the older and less precise CLAS-published results of Ref~\cite{Dugger:2005my} (green open squares with both error bars).  The three analyses used  substantially different techniques to obtain cross sections.  The methods used in the present case have been discussed above. The analysis in  Ref.~\cite{Williams:2009yj}  was based on cross section projections of an event-based maximum-likelihood fit and partial wave analysis.  The analysis in  Ref.~\cite{Dugger:2005my}  was based on a different data set using a single-track trigger, from which the $\eta^\prime$ was extracted using the missing mass off a proton and using conventional background subtraction methods.   For both comparisons the published data closest to the center of the present 100~MeV-wide bins are shown.   It is evident that the agreement is very good over the majority of the measured range, particularly at 2.35 GeV and above, where we extract cross sections for the $f_1(1285)$ meson.  Some discrepancy exists in the lowest energy bin and in the forward-most angle bins.  The smoothed differences were used to help estimate the bin-to-bin systematic uncertainty of the present results, which has been  folded into our results shown below.  Similar agreement was found when comparing cross sections extracted from the decay mode $\etapr \rightarrow \gamma\rho^0$.   Numerical results for the present differential cross section measurements of the $\eta^\prime$ are tabulated in Appendix~\ref{apx:dcs_numbers_etap}.

\section{Results}
\label{sec:results}
\subsection{Mass and width}

\begin{figure}
  \centering 
    \label{fig:mass}
  \begin{overpic}[width=0.5\textwidth,angle=0]{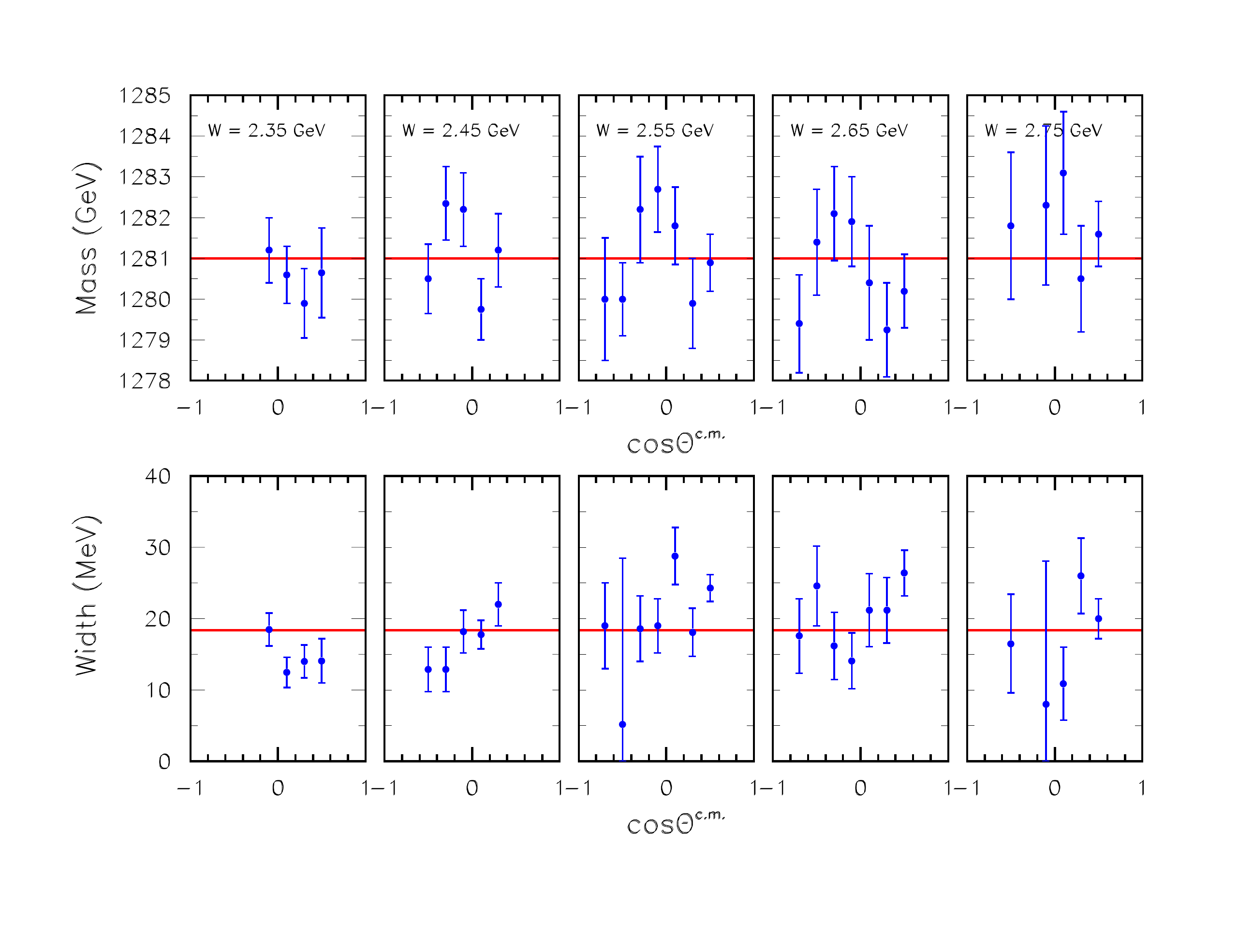}
      \put(-1,55){\large{(a)}}
      \put(-1,27){\large{(b)}}
   \end{overpic}
   \vspace{-1.0cm}
  \caption{The (a)  Mass and (b) Lorentzian width $\Gamma$ of the meson from fits binned in $W$ and $\cos\Theta^{c.m.}$. The width parameters were part of fits to a Voigtian line shape using the bin-dependent experimental mass resolution.  The overall weighted mean of each quantity is  shown by the corresponding horizontal red lines. See the summary in  Table~\ref{tab:R_Mass_and_Width_Results}.
  }
  \label{fig:Mass_and_Width}
\end{figure}

From the fits to the missing mass spectra, MM$(\gamma,p)$, for the events in the $\eta\pi^+\pi^-$ decay mode, we determined the mass and width of the initially-unidentified meson state (``$x$").  This was done in all kinematic bins in $W$ and meson production angle that had sufficiently-good signal-to-noise characteristics.  A survey of the consistency of the results is given in Fig.~\ref{fig:Mass_and_Width}.  Similar fits were made for the $\eta^\prime$ mass and  (resolution-dominated) width. Averaging these individual determinations together leads to the final values shown in Table~\ref{tab:R_Mass_and_Width_Results}, shown with the PDG values for the two listed mesons near 1285 \mmunit.  The given uncertainties are the combined systematic and statistical values.
The \etapr\ mass differs by 0.70 MeV from the world average, which is consistent with the known mass accuracy of the CLAS system.  This possible bias is included in the estimated total uncertainty of the mass of the  \xx\ meson.

\begin{table}[htb]
  \centering
  \def \mmunit {{$\mathrm{MeV/c^2}$}}  
  \begin{tabular}{ lccc }
  \hline
  \hline
  Channel & & Mass (\mmunit) & Width (MeV) \\
  \hline
  $\eta^\prime \to \eta \pi^+ \pi^-$ & CLAS & 958.48 $\pm$ 0.04 & $\Gamma \ll \sigma_{exp}$\\
  $x \to \eta \pi^+ \pi^-$     & CLAS  & 1281.0 $\pm$ 0.8  & 18.4 $\pm$ 1.4\\
  \hline
  \etapr     & PDG   & 957.78 $\pm$ 0.06  & 0.198 $\pm$ 0.009 \\
  \fx          & PDG   & 1281.9 $\pm$ 0.5   &  24.2 $\pm$ 1.1 \\
  \etax      & PDG   & 1294   $\pm$ 4     &  55   $\pm$ 5 \\
  \hline
  \hline
  \end{tabular}

  \caption{\etapr\ and \xx\ masses and Voigtian widths compared to the PDG values~\cite{Agashe:2014kda} for the \etapr, \fx, and  \etax.  The uncertainties are the combined statistical and systematic values. The width of the \etapr\ is not reported since it is much smaller than our experimental mass resolution of 3-6 MeV/$c^2$.  }
  \label{tab:R_Mass_and_Width_Results}
\end{table}

The mass of $1281.0 \pm 0.8$~MeV measured in this experiment is in very good agreement with the world average for the \fx\ state.  The uncertainty is also comparable to the previous world average for this state.  Also, the measured mass is quite incompatible with that of the \etax\ state.  The measured Lorentzian width of $\Gamma = 18.4 \pm 1.4$~MeV was obtained from the Voigtian fits of the meson, as summarized in Fig.~\ref{fig:Mass_and_Width}.  The width is about $4\sigma$ smaller than the world average of the \fx, and very much smaller than that of the \etax\ (see Table~\ref{tab:R_Mass_and_Width_Results}).  The identity of the meson seen in this experiment therefore leans strongly in favor of the well-known \fx\ and away from the less-well established \etax.

One caveat must be mentioned, however.  The Brookhaven E852 experiment, cited results~\cite{Manak:2000px} from $\pi^- p \to \eta \pi^+ \pi^- n$ showing an \etax\ with mass $1282 \pm 5$~MeV, compatible with the present  measurement. The related PWA analysis of $\pi^- p \to K^+ K^- \pi^0 n$ data~\cite{Adams:2001sk} found the width for the $1^{++}$ (\fx) wave at $\Gamma = 45 \pm 9 \pm 7$~MeV, while fitting only the intensity function for the $1^{++}$-wave yielded a much smaller width, $\Gamma = 23 \pm 5$~MeV. They concluded that interference between the \fx\ and \etax\ was significant in that reaction.  The very large background under the \fx\ signal in the present data precluded doing a full partial wave analysis of the signal region. We cannot completely rule out interference effects between the \fx\ and \etax\ influencing our observed width, but see Sec.~\ref{sec:amplitude} which shows negligible $0^-$ contribution to our data.  The width of the observed state is more compatible with the world-average results of the \fx\ than the \etax. It is also in excellent agreement with width of the \fx\ obtained using PWA in central production by the fairly recent E690 experiment~\cite{Sosa:1997qm}. In the following discussion of cross sections, it will be assumed that the only relevant contribution to the signal is from the  \fx.

\subsection{Differential Cross Sections}
\label{subsec:crosssections}

We  present the \fx\ differential photoproduction cross section into the  $\etapipi$ final state, uncorrected by the branching fraction $\Gamma(\fx \to \eta\pi\pi)/\Gamma(\fx \to all)$.  CLAS was not sensitive to all-neutral decay modes of the \fx, nor could the strong four-pion decay mode be measured with precision, hence the total rate was not measurable.   The data were binned in  $W$ and  $\cos\Theta^{c.m.}$ in the overall center-of-mass frame.   Ten 100~MeV-wide bins in $W$, from 1.8 to 2.8 GeV were defined.  Nine bins in $\cos\Theta^{c.m.}$, eight of width $0.2$ from $-0.8$ to $+0.8$, and one bin $0.1$ wide from $+0.8$ to $+0.9$. The forward and backward holes of the \clas\ detector were the limiting factors in the angular coverage.

\begin{figure*}[htb]
\centering 
  \includegraphics[width=1.0\textwidth]{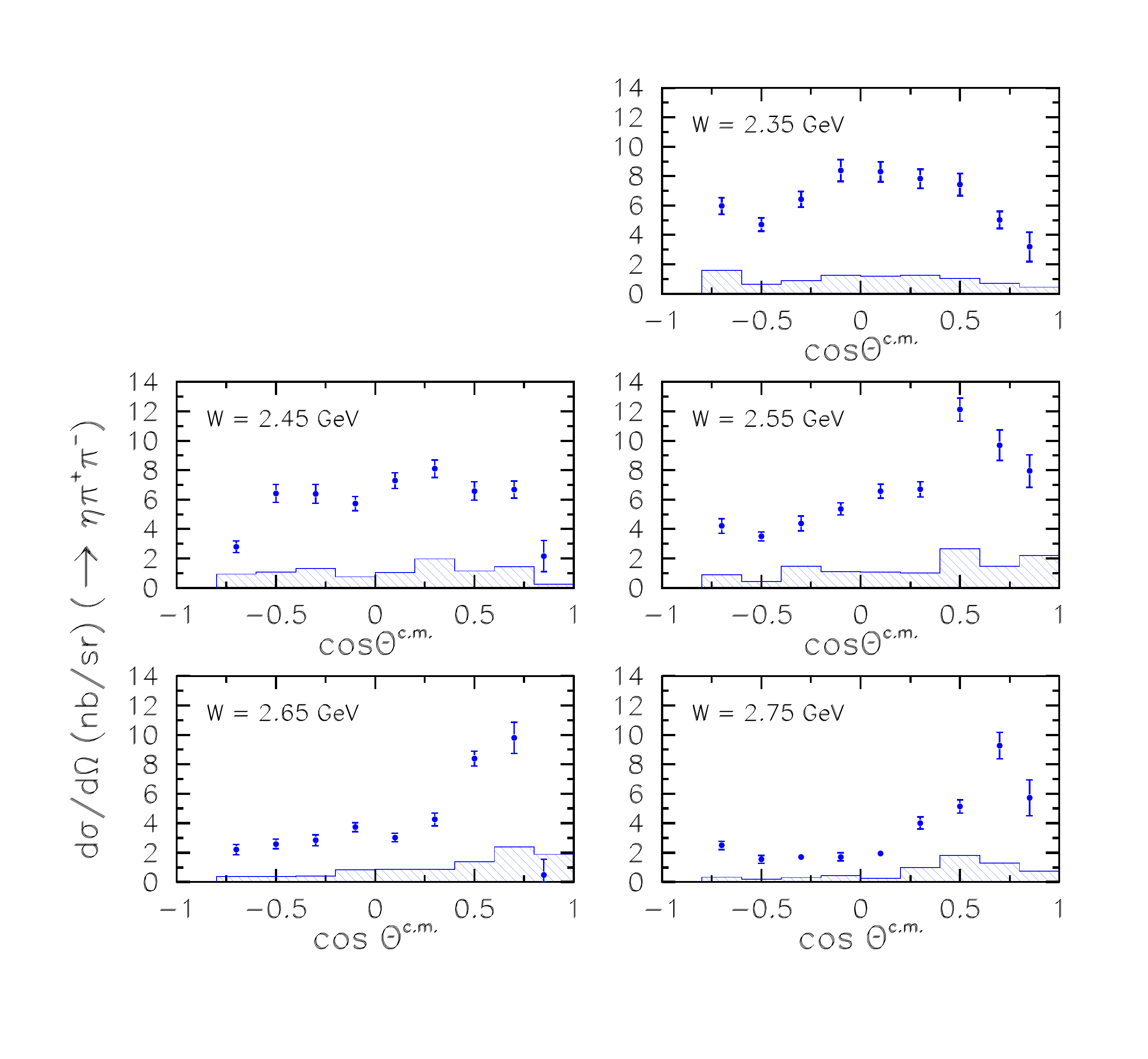}
  \centering
  \vspace{-2cm}
  \caption
 {(Color online) Differential cross section $d\sigma / d\Omega$ for $\gamma p \to
  \fx p$ with $\fx \rightarrow \eta \pi^+ \pi^-$ as a function of the meson c.m. production angle. The vertical error
  bars are the statistical uncertainties from fitting and from combining two decay modes. The shaded histogram is the total
  point-to-point systematic uncertainty discussed in the text. Each  panel shows a 100~MeV wide bin centered at the indicated $W$.}
  \label{fig:R_dcs_domega_x1280_etapipi_kkpi_combined}
\end{figure*}

The results are a combination of three analyses: $\eta\pi^+\pi^-$ data extracted with a Voigtian fit for event yields, the same $\eta\pi^+\pi^-$ events fit to a combination of Monte Carlo signal plus several simulated multi-pion backgrounds, and $K^\pm K^0 \pi^\mp$ data fit with a Voigtian signal and polynomial background function. The cross sections extracted in $K^\pm K^0 \pi^\mp$  have been scaled by the measured (in this experiment) branching ratio
\begin{equation}
  \frac{\Gamma( \fx \rightarrow K\bar{K}\pi)}{\Gamma( \fx \rightarrow \eta\pi\pi)}
\end{equation}
before taking the weighted mean of the independently-extracted measurements in these two decay modes.  The branching ratio result will be presented in Sec.~\ref{subsec:branching_fractions}. The measured differential cross sections presented in Fig.~\ref{fig:R_dcs_domega_x1280_etapipi_kkpi_combined} are thus for the decay to $\eta\pi^+\pi^-$, but with the event statistics from the  kaonic decay modes included.  The total systematic uncertainties shown include both the values from Table~\ref{tab:final_systematics} and from the yield extraction methods in the \etapipi\ channel discussed previously.   We estimate the overall systematic uncertainty to be between 11\% to 14\% for the differential cross sections.

The differential cross section shows some structure in production angle already in the near-threshold bin for $2.30<W<2.40$~GeV  (threshold is at 2.22~GeV).  The cross section falls off in the forward-most angle bins, which is not typical in meson photoproduction.  For $W$ above 2.55 GeV, a forward rise becomes more pronounced, although the drop in cross section at very forward angles persists.  A forward-angle rise is usually associated with $t$-channel processes, while the fall back towards zero in the extreme forward direction can occur either with multiple Regge-exchange trajectories or in the analytic structure of a single $t$-channel amplitude.

\begin{figure}[htb]
  \centering
  \vspace{-5mm}
  \hspace{-5mm}
  \includegraphics[width=0.5\textwidth]{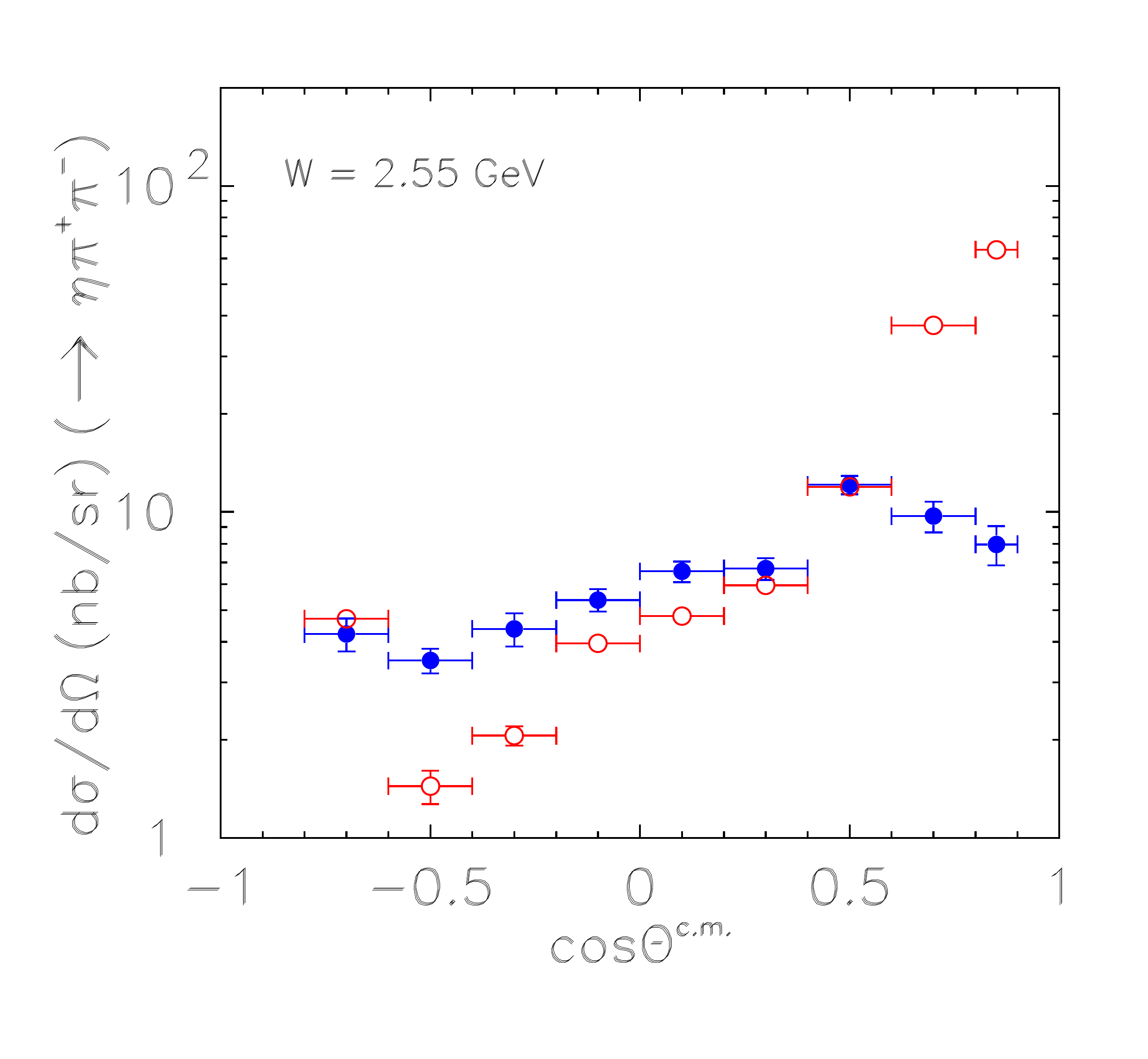} 
  \vspace{-10mm}
  \caption []{ (Color online) Cross section comparison for  $\gamma p \to \fx p \to \eta \pi^+ \pi^- p$ (blue full circles)   compared to  $\gamma p \to \etapr p \to \eta \pi^+ 
  \pi^- p$ (red open circles) at $W = 2.55\;{\rm  GeV}$. The \fx\ differential cross section is not  forward-peaked like the \etapr. Note the logarithmic vertical scale.}  
  \label{fig:dcs_x1280_etaprime_comparison}
\end{figure}

Figure~\ref{fig:dcs_x1280_etaprime_comparison} shows a cross section comparison for \etapr\ and $\fx \to \etapipi$ at $W = 2.55\;{\rm GeV}$.  The \etapr\ cross section exhibits much stronger $t-$ and $u-$channel signatures in its $\cos \Theta^{c.m.}$ dependence than does the \fx, which is quite ``flat'' by comparison.  The same is true in all measured $W$ bins, and also true if the comparison is made at equal excess energy above the respective reaction thresholds.  This may imply that the \fx\ photoproduction mechanism is less peripheral than that of the \etapr,  not dominated by $t$-channel production processes.

\begin{figure*}[htb]
  \includegraphics[width=1.0\textwidth]{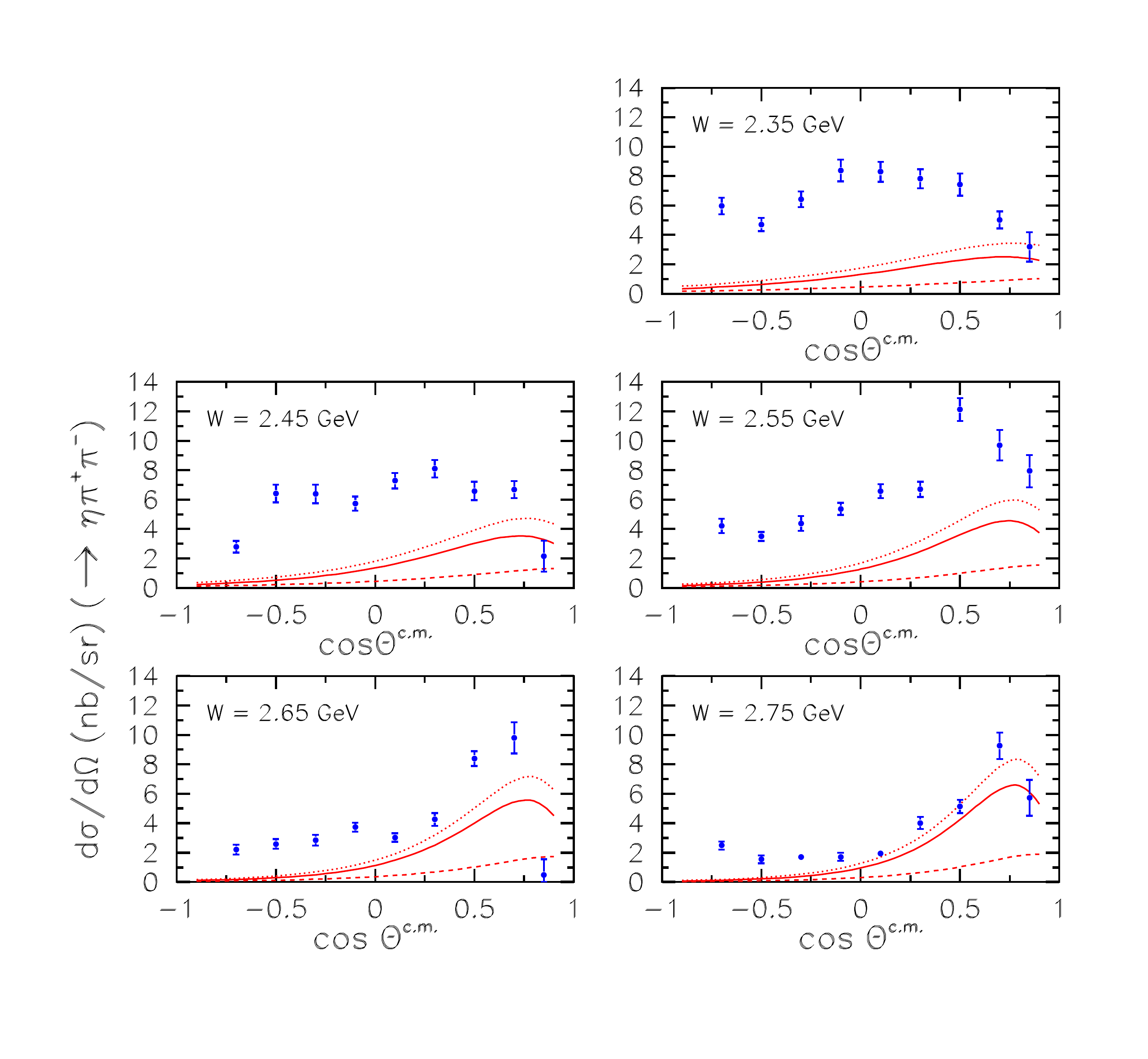}
  \centering
  \vspace{-2cm}
  \caption{(Color online) Differential cross sections $d\sigma / d\Omega$ for
    $\gamma p \to \fx p$ and
    $\fx\ \rightarrow \eta \pi^+ \pi^-$ as a function of the meson c.m. production angle for five values of the frame-invariant energy, $W$. The error bars are statistical only.  The solid red line is the prediction by Kochelev~\cite{Kochelev:PhysRevC.80.025201} for the \fx\ and the dashed red line is the corresponding prediction for the \etax. The curves have been scaled by the PDG branching fraction for $\fx \to \eta\pi^+\pi^-$ (see text).  The dash-dot line is the incoherent sum of both mesons.}    
  \label{fig:R_dcs_domega_x1280_kochelev}
\end{figure*}

The cross sections can be compared to Regge-model predictions by Kochelev~\cite{Kochelev:PhysRevC.80.025201} for both the \fx\ and \etax\ states.  The model calculations were recomputed~\cite{Kochelev:private}  for our choice of energy and $\cos \Theta^{c.m.}$ bins and are shown as $d\sigma / d\Omega$ in Fig.~\ref{fig:R_dcs_domega_x1280_kochelev}.  The curves show predictions for both the \fx\ and \etax\ and for their incoherent sum. All model curves have been scaled by the PDG branching fraction $\Gamma(\fx \to \eta\pi^+\pi^-)$ for this comparison:  $0.52\times(2/3)$.   This is an ad-hoc scaling for the poorly-known \etax, which has been observed in $K\bar{K}\pi$ final states~\cite{Adams:2001sk, Ahohe:2005ug}, but not in the  \etapipi\  final state.

The Kochelev prediction utilizes $t$-channel meson production, with the exchange of  $\rho$ and $\omega$ trajectories.  The model uses phenomenological couplings from related reactions with vector-meson-dominance inspired hadronic from factors, and was adjusted to match the well-known pseudoscalar states $\eta$ and $\eta^{\prime}$.  In comparison, the present results show clearly that the $t$-channel alone does not reproduce our measurements, especially near threshold. In the highest-energy bins the  \fx\ model converges towards the data points in the forward production angle bins, but the middle and backward angles are not reproduced by the model. Our interpretation of Kochelev's model and the comparison to the \etapr\ cross sections (Fig.~\ref{fig:dcs_x1280_etaprime_comparison}) suggest that part of the strength of \fx\ production comes from $s$-channel processes.  That is, the decay of excited $N^*$ intermediate baryon states may be important here.

\begin{figure*}[htb]
  \centering
  \includegraphics[width=1.0\textwidth]{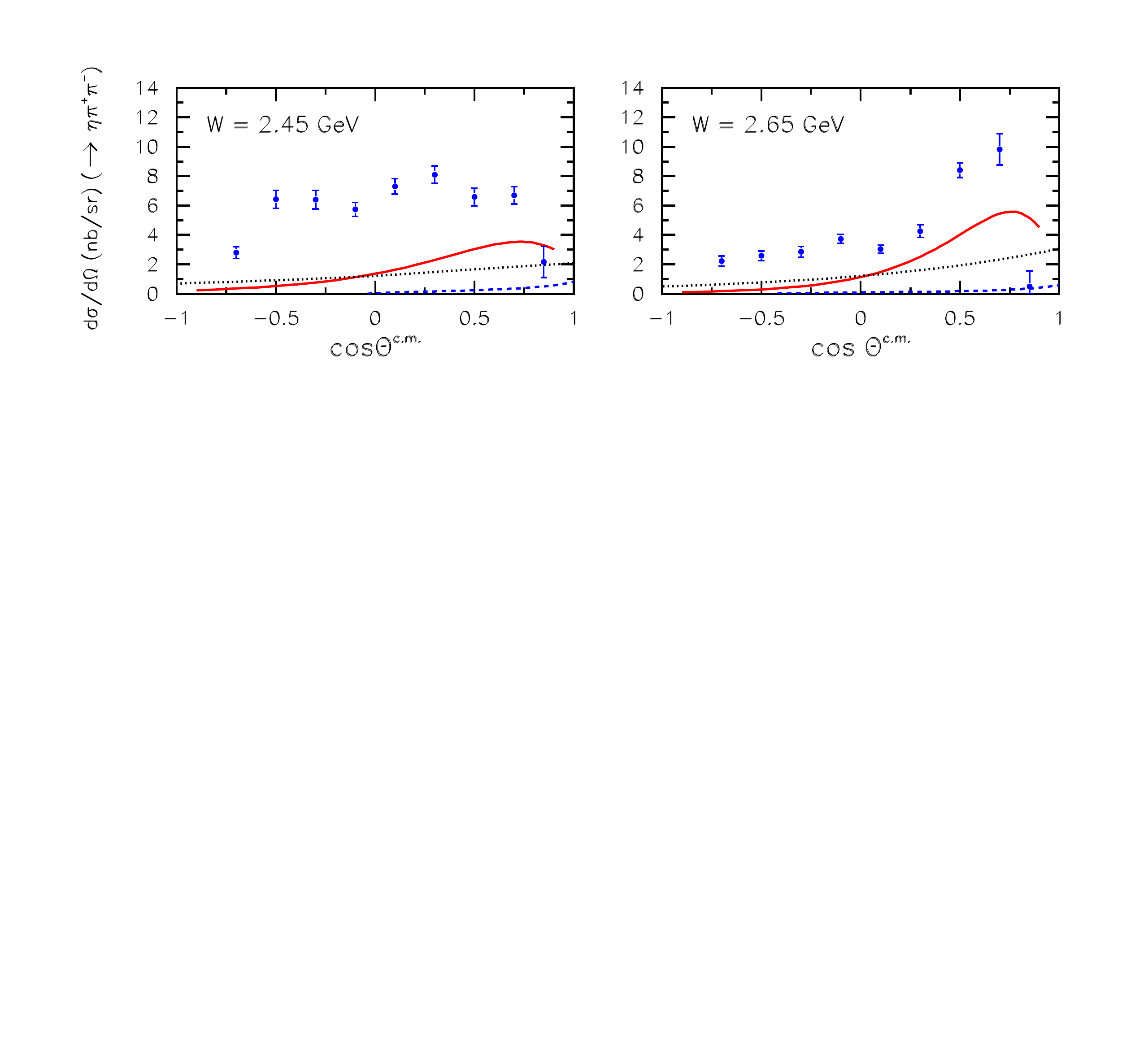}
  \vspace{-11cm}
   \caption{(Color online) Predictions from several models of \fx\ photoproduction compared  to the present results at  $W=2.45$~(GeV) and $W=2.65$~GeV. The Kochelev~{\it  et al.}~model prediction \cite{Kochelev:PhysRevC.80.025201}  for the \fx\ (solid red) is generally closest to the data, but the match is poor.  The model of Domokos~{\it et al.}~\cite{PhysRevD.80.115018} (dashed blue) is based on $\rho$ and $\omega$ exchange using a specific model for the coupling to the \fx\.. The meson-exchange model  prediction of Huang {\it et al.}~\cite{Huang:2013jda, XiePrivComm} (dotted black) used hadronic form factors that are not constrained by other reactions.
    }
  \label{fig:3models}
\end{figure*}

Figure~\ref{fig:3models} shows a comparison of three models for the \fx\ in energy bins for $W = 2.45$ and $2.65$~GeV.  Apart from the  Kochelev {\it et al.} model (in red), a model based upon a different theoretical starting point was published by S.~Domokos~{\it et  al.}~\cite{PhysRevD.80.115018} (dashed blue).  The model was motivated by Chern-Simons-term induced interactions in holographic QCD. It calculates anomalous couplings that link vector and axial-vector photoproduction, as derived from the general principles of AdS/QCD (anti-de Sitter) correspondence.  The \fx\ was stated to be an especially ``clean`` example where the model could be applied.  Calculations using single-particle $\rho$  and $\omega$ exchange at low $s$  were presented, as well as a separate Reggeized meson exchange picture for large $s$ and small $|t|$.   We plot the $(\rho, \omega)$-exchange version of their model at our kinematics after checking that we could reproduce the Reggeon-exchange calculation that is plotted in their paper. The curves are scaled by the PDG decay branching fraction for $f_1(1285) \to \eta \pi^+ \pi^-$. The prediction is much smaller than the new data, even in the forward-most region where the $t-$channel process is dominant,  the kinematic region where this prediction was claimed  to be most characteristic of \fx\ photoproduction.  The Reggeized version of the calculation is not shown since it was even more incompatible with the results.  

A third model, by Huang~{\it et al.}~\cite{Huang:2013jda, XiePrivComm} (dotted black), uses an effective-Lagrangian approach with tree-level $\rho$ and $\omega$ exchange.  The cross section magnitudes depend very sensitively on the hadronic form-factor cutoffs at the $f_1 V \gamma$ and $VNN$ vertices  ($V=\rho,\omega$), and were adjusted without theoretical  linkage to other reaction channels.  The curves shown in Fig.~\ref{fig:3models} are for 1.0 GeV cutoffs, and scaled by the PDG decay branching fraction for $f_1(1285) \to \eta \pi^+ \pi^-$.  Again, for this model the data are not well reproduced by the prediction made by this calculation.  We can conclude that none of these three model predictions are close to the mark in describing the reaction mechanism leading to \fx\ photoproduction.  This suggests that  $s$- and $u$-channel mechanisms ($N^*$ decay and baryon exchange), or perhaps a non-$\bar{q}q$ structure of the \fx~\cite{Aceti:2015pma}\cite{Geng:2015yta} may need to be considered.

Numerical results for the measured differential cross sections are tabulated in Appendix~\ref{apx:dcs_numbers}.

\subsection{Branching Ratios}
\label{subsec:branching_fractions}

The experiment did not measure the four-pion decay modes of the \fx\ and hence could not determine  absolute branching fractions.  Instead, we measured the ratios of acceptance-corrected yields in the \etapipi, \kkpi, and \rhogamma\ decay modes.  The branching ratios measured in this analysis are 
\begin{align}
  &\frac{\Gamma( \fx  \rightarrow K\bar{K}\pi)}
        {\Gamma( \fx \rightarrow \eta\pi\pi)}\quad \mathrm{and}
  \label{eq:br_kkpi_etapipi} \\ 
  &\frac{\Gamma( \fx \rightarrow \gamma\rho^0)}
        {\Gamma( \fx  \rightarrow  \eta\pi\pi)}. 
  \label{eq:br_rhogamma_etapipi}
\end{align}

\begin{table*}[htb]
  \centering

\providecommand{\e}[1]{\ensuremath{\times 10^{#1}}}

\begin{tabular}{lcccc}
  \hline
  \hline
  Item & Value & Statistical    & Systematic  & PDG  \\
          &           & Uncertainty & Uncertainty & $f_1(1285)$ \\
  \hline
  $\eta\pi^{+}\pi^{-}$ Event Yield & 1.33\e{5} & 4.9\e{3} &  2.9\e{3} &\\
  $\eta\pi^{+}\pi^{-}$ Acceptance  & 0.0652 & 9.7\e{-5} & 0.0072 &\\
  \hline
  $K^{\pm}K^0\pi^{\mp}$ Event Yield  & 6570   &   180 &   340 &\\
  $K^{\pm}K^0\pi^{\mp}$ Acceptance & 0.0149 & 3.18\e{-5} & 0.0016 &\\
  $\gamma\rho^{0}$ Event Yield & 3790 &   790        &  850 &\\
  $\gamma\rho^{0}$ Acceptance  & 0.0248    & 6.4\e{-5} & 0.0050 &\\
  Isospin C.G. $\Gamma(K^{\pm}K^0\pi^{\mp}) / \Gamma(K\bar{K}\pi)$ & 2/3 & & &\\
  Isospin C.G. $\Gamma(\eta\pi^{+}\pi^{-}) / \Gamma(\eta\pi\pi)$  & 2/3 & & &\\   
  $\gamma\rho^{0}$ correction from $\eta'$ $d\sigma/d\Omega$ & 0.95  & & &\\
  \hline
  Branching Ratio $\Gamma(K \bar{K}\pi)  /  \Gamma(\eta\pi\pi)$  & 0.216 & 0.010 & 0.031 & 0.171 $\pm$ 0.013 \\
  Branching Ratio $\Gamma(\gamma\rho^{0}) / \Gamma(\eta\pi\pi)$  & 0.047 & 0.010 & 0.015 & 0.105 $\pm$ 0.022  \\
    Branching Ratio $\Gamma(a_0\pi \text{\small{ (no }} \bar{K} K\text{\small{)}}) / \Gamma(\eta\pi\pi \text{\small{ (all)}}) $  & 0.74 & 0.02 & 0.09 & 0.69 $\pm$ 0.13  \\
\hline
  \hline
\end{tabular}
\caption[]{Branching ratios of the \fx\ meson, with estimated uncertainties from all sources.}
  \label{tab:R_Branching_Ratio}
\end{table*}

In computing these ratios there were several possible ways to combine the particle yields. In the $\eta\pi^+\pi^-$ and $K^\pm K^0 \pi^\mp$ decay modes there were sufficient statistics to compute particle yields for each kinematic bin used to compute the differential cross sections. Summing the partial yields then determined the total yield for each decay channel.  This method used the bin-dependent widths ($\sigma$) in the Voigtian function to parametrize the experimental resolution, rather than using a single global value.  Also, the results of our systematic studies of the particle yield and  acceptance could be applied.  An alternative method was to fit the missing-mass spectra integrated over $W$ and $\cos\Theta^{c.m.}$, excluding bins with insufficient acceptance. This was the only possible method for computing the radiative decay ratio, because the \fx\ signal in the $\gamma\rho^0$ decay mode was quite small.  The systematic uncertainty on the acceptance of $\fx \to \gamma\rho^0$ events was estimated by iterating the Monte Carlo to match the observed differential cross sections.  Table~\ref{tab:R_Branching_Ratio} summarizes this information and shows comparisons to world data for the \fx. Isospin Clebsch-Gordan factors were applied to each decay mode.  No branching fractions have been reported for the \etax, even though this state has been observed in $K\bar{K}\pi$ final states with strength comparable to the \fx~\cite{Adams:2001sk}.

We find our value for the $K$-decay ratio, Eq.~\ref{eq:br_kkpi_etapipi}, to be larger than the PDG value for the \fx, but consistent within the measured uncertainties.  The radiative decay ratio however, Eq.~\ref{eq:br_rhogamma_etapipi}, is lower than the world average by a considerable amount.  We find a ratio of $0.047 \pm 0.018$ as shown in Table~\ref{tab:R_Branching_Ratio}, which is less than half the PDG average value of $0.105 \pm 0.022$. Even with our large uncertainty in the \rhogamma\ yield extraction, we find a roughly ``3$\sigma$'' difference between our value the PDG fit.

The radiative decay ratio ${\Gamma(x \to \gamma\rho^0)}/{\Gamma(x \to \eta\pi\pi)}$ is interesting because there are both experimental and theoretical values for comparison.  Table~\ref{tab:rad_decay_ratios} lists the calculated widths of both the \fx\ and \etax\ mesons from several models compared to the present work combined with PDG information. The CLAS measurements of the total width $\Gamma$ (in MeV) and the measured radiative  branching ratio ${\Gamma(x \to \gamma\rho^0)}/{\Gamma(x  \to   \eta\pi\pi)}$ can be combined with the PDG value for ${\cal   B}(\fx \to \etapipi)$ to compute the width for the radiative decay. On the other hand, one can take the PDG total width and the PDG radiative decay branching fraction and again compute the expected radiative width.  The result from the present work of $453\pm177$~keV is in poor agreement with the PDG-based estimate of $1331\pm320$~keV. Our estimate  is not quite independent of all previous work since the branching fraction to $\eta\pi\pi$ is assumed to be accurate.

\begin{table*}[htbp]
  \centering
  \begin{tabular}{ ccc }
    \hline \hline
    Theory   & Prediction                                                                                     & $\Gamma(\gamma\rho^0)$\\ 
    \hline
    \multirow{4}{*}{Lakhina and Swanson~\cite{Swanson}}   & Relativistic \fx    & 480 keV  \\ &
                                                                                          Non-Rel. \fx              & 1200 keV \\ & 
                                                                                          Relativistic \etax       & 240 keV  \\ &
                                                                                          Non-Rel. \etax          & 400 keV  \\ 

    \multirow{2}{*}{Ishida {\it et al.}~\cite{PhysRevD.40.1497}} &\fx \quad$\Theta_1$ & 509 keV \\ &
                                                                  \fx \quad$\Theta_2$ & 565 keV \\
    \hline \hline 
    Experiment & $\Gamma \times {\cal B}(\fx \to \eta\pi\pi)_{{\rm PDG}} \times   (\Gamma(\gamma\rho^0) / \Gamma(\eta\pi\pi))$           &    \\ 

    \hline 
    CLAS & $(18.4 \pm 1.4\;{\rm MeV}) \times (0.524 \pm .002)   \times (0.047 \pm 0.018) $     &    $ 453 \pm 177\;{\rm keV} $ \\ 

    PDG \fx ~\cite{Agashe:2014kda} & $(24.2 \pm 1.1\;{\rm MeV}) \times (0.055 \pm 0.013) $&$   1331 \pm 320 \;{\rm keV} $ \\ 
    \hline  \hline 
  \end{tabular}
  \caption{Predictions for radiative decay widths $x \to \gamma\rho^0$   for two models,  compared to the CLAS-measured results using the total width, $\Gamma$, and branching ratio  $\Gamma (\gamma \rho^0) / \Gamma(\eta \pi \pi)$.  Alternative comparison is made to the current PDG estimate.}
  \label{tab:rad_decay_ratios}
\end{table*}


For comparison, we have quark model radiative decay predictions by O.~Lakhina and E.~Swanson~\cite{Swanson}.  These use a non-relativistic Coulomb-plus-linear quark potential model and predict a $\Gamma({\fx \to \rhogamma})$ of 480~keV in a relativized version of the calculation, while the non-relativistic version predicts 1200~keV. One sees that the present results are in better agreement with the relativized version of this model, while the PDG-based estimate would favor the non-relativistic result.  Lakhina and Swanson also calculated values for $\Gamma(\etax \to \rhogamma)$ of 240~keV and 400~keV.  The axial-vector \fx\ is predicted to have a stronger coupling $g_{\gamma \rho x}$ than the pseudoscalar \etax.  But without a corresponding value of $\etax \to \etapipi$ from either experiment or theory, these values can not be compared to our experimental ratio.

Another  prediction for the \fx\ width was made by S.~Ishida {\it et~al}. using a covariant oscillator model~\cite{PhysRevD.40.1497}.  It predicted a radiative width for $\fx \to \gamma\rho^0$ of between 509 to 565~keV depending on a particular mixing angle.  This prediction is narrower than the PDG-based estimate, but consistent with our experimental result.

In summary, the two branching ratios presented here are only partly consistent with previous experimental results for the \fx.  Since nothing is known from experiment about  \etax\ branching fractions, nothing more conclusive can be said about the  identity of the observed  meson from the CLAS branching ratio  results.  Model calculations are at present not decisive in this regard, either.

\subsection{Dalitz Analysis of the Decay to $\eta\pi\pi$}
\label{sec:dalitz_analysis}

The three-body decay of the observed meson ``$x$" to $\eta\pi^+\pi^-$ can be examined for evidence of its intrinsic spin and its decay substructure.  We presume initially to have no knowledge of the meson's identity.  The identity of the meson would be confirmed if the spin were shown to be either 0 (pseudoscalar \etax) or 1 (axial-vector \fx).  Amplitude analysis of a Dalitz distribution~\cite{Dalitz:1953cp} is a well-established tool for investigating the dynamics of any three-body decay. The partial decay rate of $m_x$ in a three-body process averaged over  spin states can be expressed in terms of the invariant masses $m_{\eta\pi^{+}}^2$ and $m_{\eta\pi^{-}}^2$ as
\begin{equation}
d\Gamma = \frac{1}{(2\pi)^3}\frac{1}{32m_x}|\overline{\mathcal{M}^2}|dm^2_{\eta\pi^+}dm^2_{\eta\pi^-}.
\end{equation}
If the meson decays into the three daughter particles with the matrix element magnitude $|\mathcal{M}|^2$ constant, then the distribution on the Dalitz plot will be uniform, filling ``phase space". However, if it decays via an intermediate resonant process, the Dalitz plot will show a non-uniform distribution, with interfering band(s) at the masses of any intermediate resonances.  The intensity distribution is determined by the angular momentum of the decay channels and the interferences among their amplitudes.

\begin{figure*}[htb]
 \subfloat{
    \label{fig:uncorrected_dalitz} 
   \begin{overpic}[width=0.48\textwidth,height=0.35\textheight,tics=10] {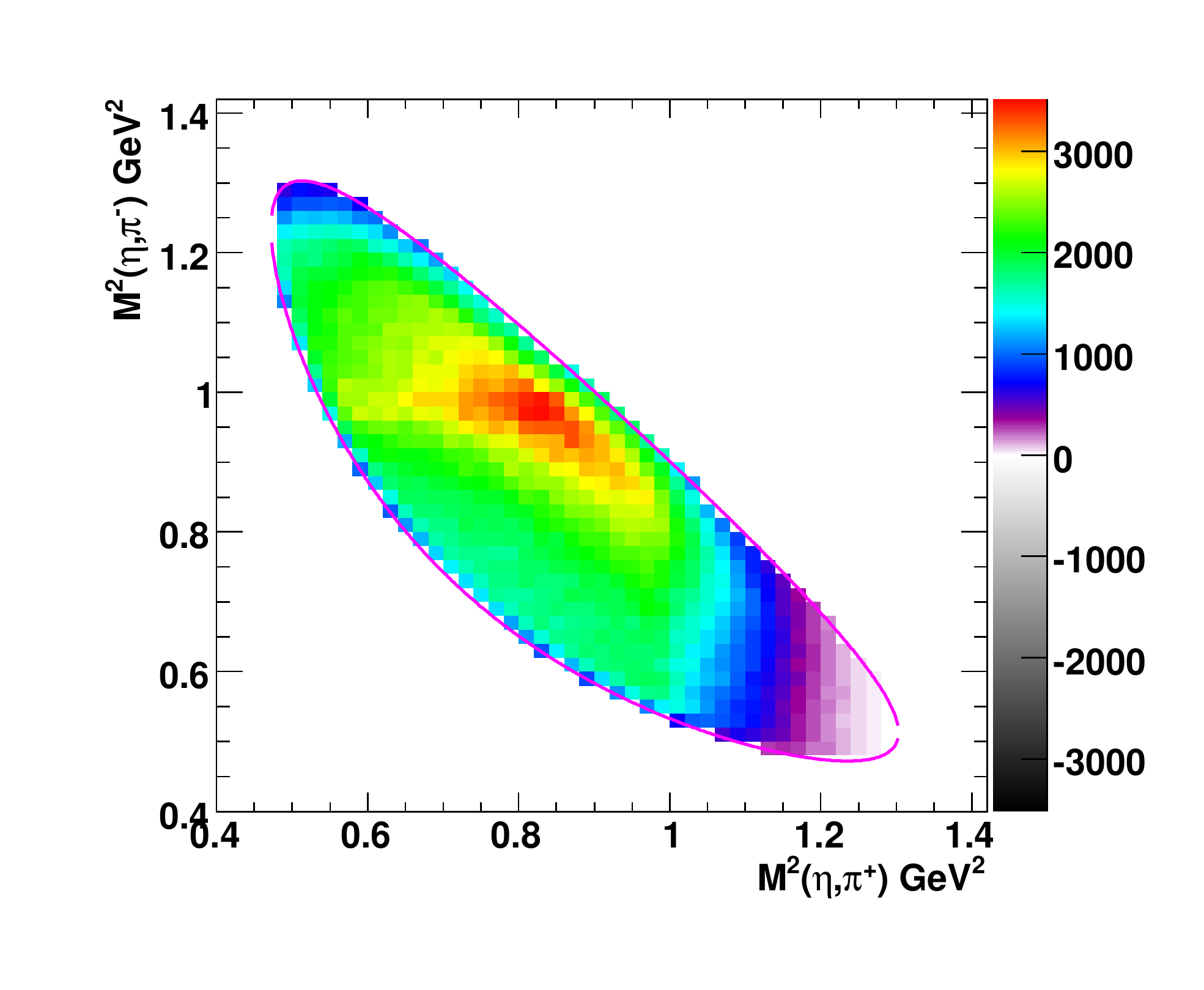}
      \put(65,75){(a)} 
   \end{overpic} 
   \label{fig:dalitz_bkgdsubtracted} 
  }
  \subfloat{
   \begin{overpic}[width=0.48\textwidth,height=0.35\textheight,tics=10] {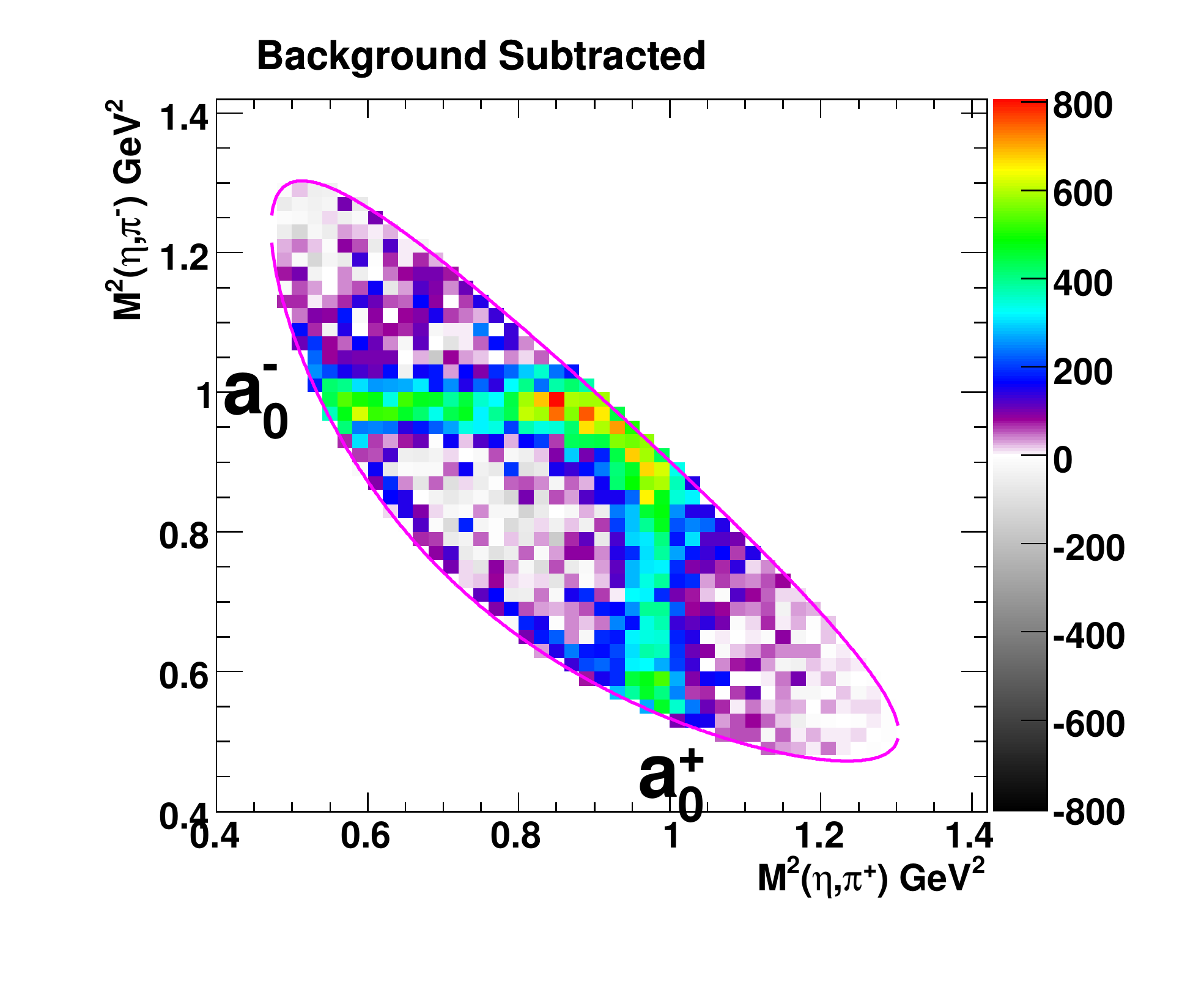}
      \put(65,75){(b)} 
   \end{overpic} 
   \label{fig:dalitz_bkgdsubtracted} 
  }
  \caption[]{(Color online) (a) Initial Dalitz plot for $\eta \pi^+ \pi^-$ events with \mmp\ between 1251 and 1311 \mmunit, prior to sideband subtraction. (b) After subtracting weighted and scaled multi-pion sidebands. The result is not yet corrected for acceptance.}
  \label{fig:uncorrected_dalitz_both}
\end{figure*}

All $\eta\pi^+\pi^-$ events with missing mass off the proton between 1251 and 1311  \mmunit\ were selected. Figure~\ref{fig:uncorrected_dalitz} shows the initial Dalitz plot for  these events.  The dominant multi-pion background hides all evidence  of resonant sub-structure in the decay of the meson since the signal-to-noise ratio is estimated at \about5\% from fits to the MM$(\gamma,p)$ spectrum.  The CLAS system acceptance was lowest near the high $\eta\pi^+$ masses, which required acceptance of low momentum negative pions.  These pions bent inward, toward the beam pipe, resulting in the  lowest particle acceptance in the experiment.

\begin{figure}[htb]
  \hspace{-20mm}
  \begin{overpic}[width=0.45\textwidth,tics=10] {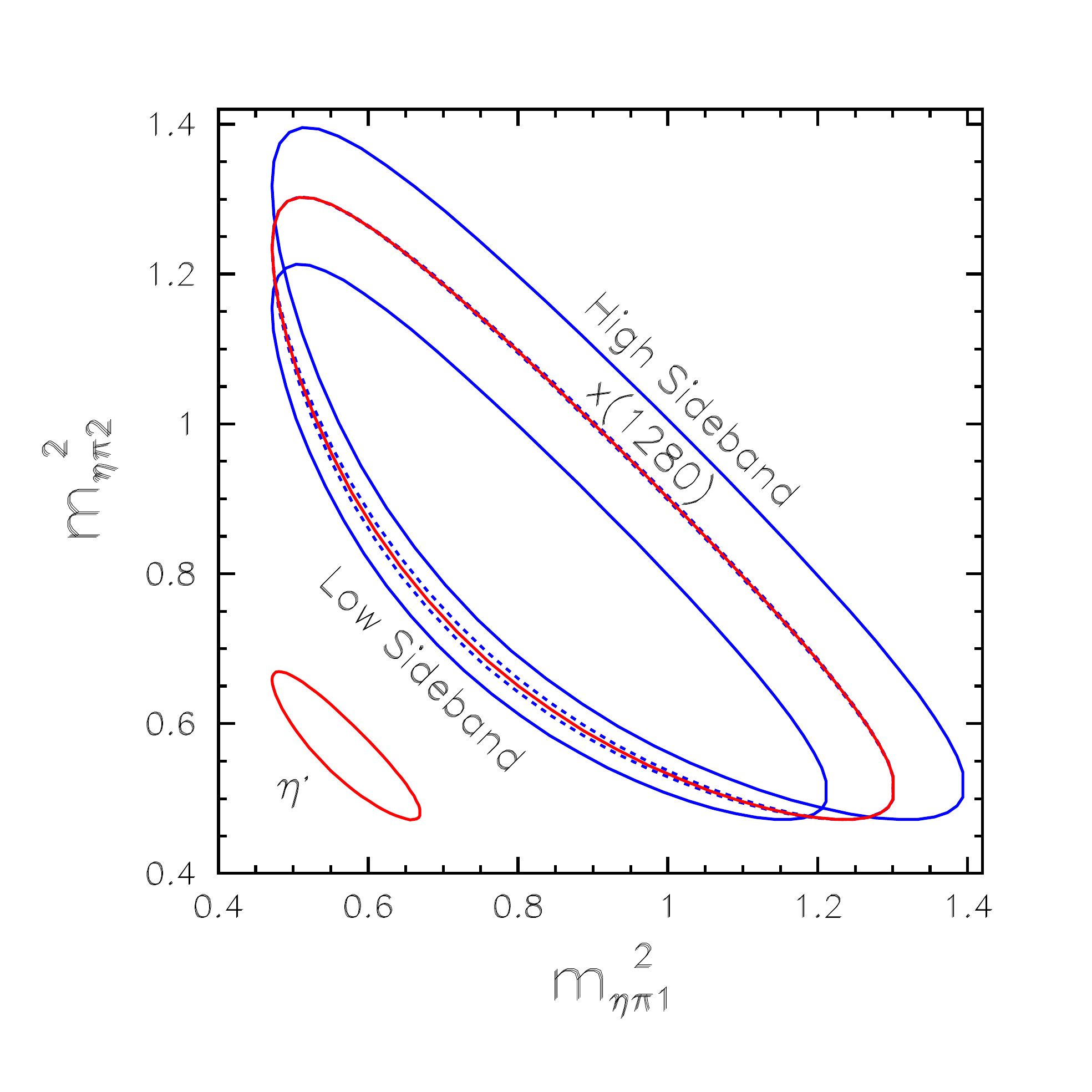}
    \put(64,64){\includegraphics[width=0.25\textwidth]{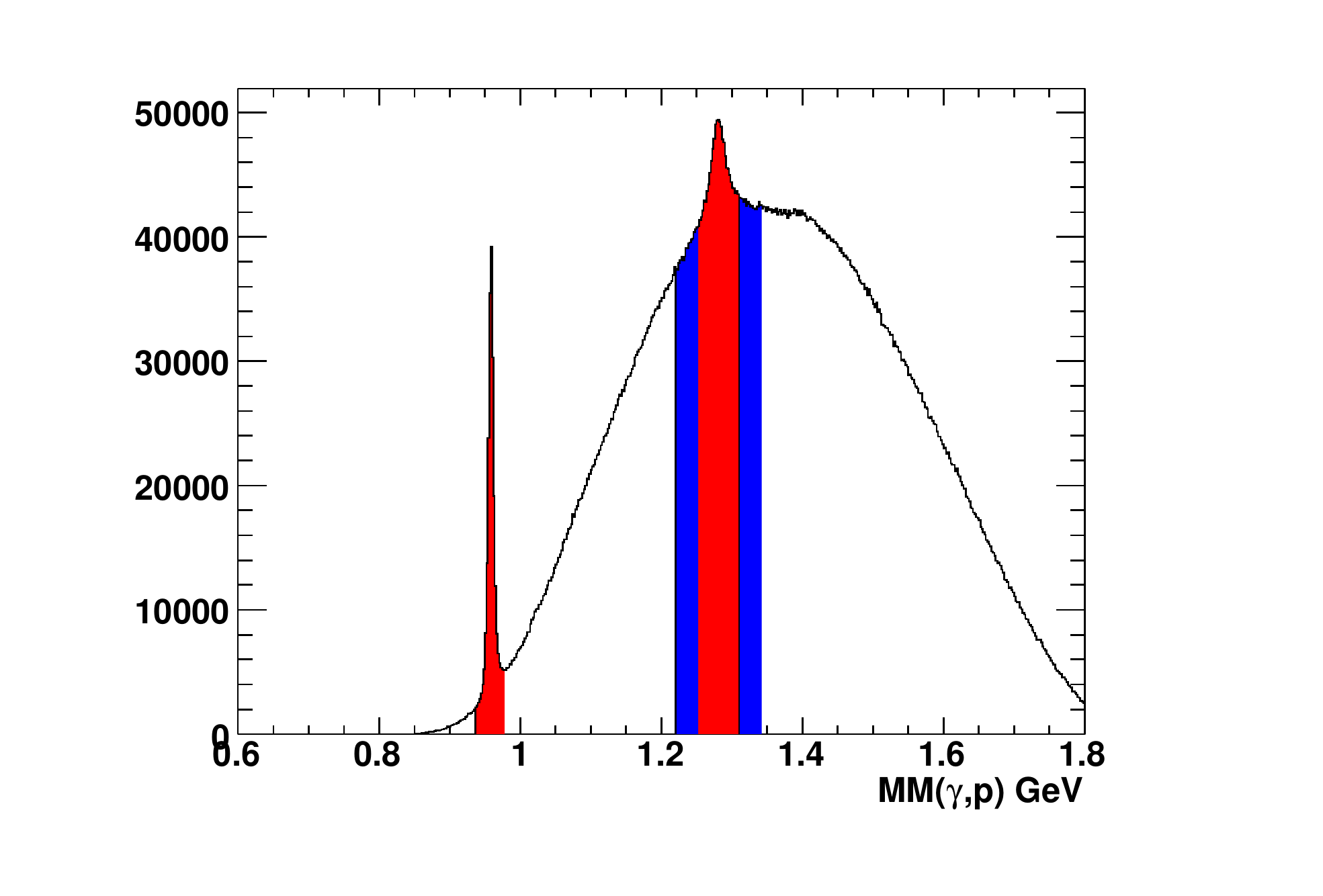}}
  \end{overpic}
  \vspace{-5mm}
  \caption{(Color online) Dalitz plot boundaries for several decays  to $\eta \pi^+ \pi^-$. The solid red curves show the limits for decay of the \etapr\ and an $x(1280)$ meson. The solid blue curves mark the limits for events with \mmp\ in 30-wide sidebands of the meson signal shown in the inset. The dashed blue curves show the transformed boundaries of sideband events using the method discussed in the text.}
  \label{fig:dalitz_bounds}
\end{figure}

The kinematic coverage of a Dalitz distribution is determined by the masses of the decaying parent particles. Figure~\ref{fig:dalitz_bounds} shows the kinematic  boundaries for the \etapr\ and \xx\ mesons and the centers of the sidebands of the \xx. The sideband overlap is far from perfect, so an alternative method was needed to remove background from this Dalitz  distribution.

Assuming the background is dominated by multi-pion events with no resonant structure, we apply a linear transformation to each invariant mass combination in the sidebands (denoted $m_{12}$ for $\pi^+\eta $ and $m_{23}$ for $\eta\pi^-$) to rescale to mass values within the signal region.  Since a phase space decay results in a flat distribution, this transformation on the $m_{12}^{2}$ and $m_{23}^{2}$ values of the sidebands should preserve the background shape and allow for accurate sideband subtraction.

The sideband-event masses $m_{ij}^2$ are rescaled to be within the signal-region boundary ${m_{ij}^\prime}^2$ according to the linear transformation
\begin{equation}
  m^{\prime\,2}_{ij} = s \left(m^2_{ij} - ({m_{ij}}^2)_{\mathbf{min}}\right) + ({m_{ij}}^2)_{\mathbf{min}},
\label{eq:dalitz_sideband_transformation}
\end{equation}
with scale-factor $s$. The bounding contour of the Dalitz plot for decay of the parent \xx\ meson is specified by limits $(m_{ij}^2)_{\mathbf{max}}$ and
$(m_{ij}^2)_{\mathbf{min}}$, and therefore by the  range
\begin{equation}
  \Delta m^2_{ij} = (m^2_{ij})_{\mathbf{max}} - (m^2_{ij})_{\mathbf{min}}.
\end{equation}
The scale, $s$, is given by
\begin{align}
  s &= \frac{\Delta m^{\prime \, 2}_{ij}}{\Delta m^2_{ij}} \\
    &= \frac{(m^\prime_x - m_k)^2 - (m_i + m_j)^2}{(m_x - m_k)^2 - (m_i + m_j)^2},
\end{align}
where $m_k$ is the mass of the third particle in the decay.  $m_x$ is the central mass of the signal region and  the shifted mass $m_x^{\prime}$ is 
\begin{equation}
  m'_x = m_x + d,
\end{equation}
where $d$ is the mass difference between the center of the signal region and the center of the sideband region.   The transformation is applied event by event, separately in both $m_{\eta\pi^+}^2$ and $m_{\eta\pi^-}^2$. The dotted lines in Figure~\ref{fig:dalitz_bounds} show that the scaled sideband regions overlay the \xx\ meson  kinematic region quite well.  

Finally, before combining the two transformed sidebands for subtraction from the signal region, they were weighted according to the estimated background in missing mass off the proton.  The spectrum in the inset of Fig.~\ref{fig:dalitz_bounds}  was fit with a Voigtian shape for the \xx\ meson and a fifth-order polynomial for the background. The transformed sideband Dalitz plots were filled according to the weight
\begin{equation}
 w = \frac{B(m_x)}{B(m^\prime_x)},
\end{equation} 
where $B(m)$ is the background polynomial evaluated at a given missing mass $m_x$. This compensates for the rising slope of the background and slightly weights the high sideband more heavily, as seen in the Figure~\ref{fig:dalitz_bounds} inset.  Monte Carlo simulations of this method verified that it does not introduce a ``bias'' in a uniformly-populated decay distribution.

Figure~\ref{fig:dalitz_bkgdsubtracted} shows the result of subtracting the scaled and weighted sideband events from those in the \xx\ meson signal region. The $a_0^{\pm}(980)$ can been seen quite clearly as bands in the resultant $\eta \pi^+ \pi^-$ Dalitz plot. The negative-count bins present are consistent with the counting statistics of the subtraction.  Thus, it appears that a substantial portion of the three body decay to $\eta\pi^+\pi^-$ goes through the two-body modes $x \to a_0^{\pm}\pi^{\mp}$ with subsequent decay of the $a_0$ to $\eta\pi$.

The CLAS acceptance for signal events was computed starting from a ``flat'' Monte Carlo distribution with the measured $\Gamma \sim 18$ MeV meson width.  Smearing due to detector resolution was accounted for by GSIM, as discussed in Sec.~\ref{sec:acceptance}. The sharp kinematic boundary in the Dalitz plot is actually ``soft'' due to the finite width of the meson  and the detector resolution. The acceptance calculation inevitably suffered from low statistics at these edges.  The analysis was therefore truncated at the boundary defined by the  centroid of the meson signal region.

\begin{figure}[htb]
  \centering
  \includegraphics[width=0.50\textwidth]{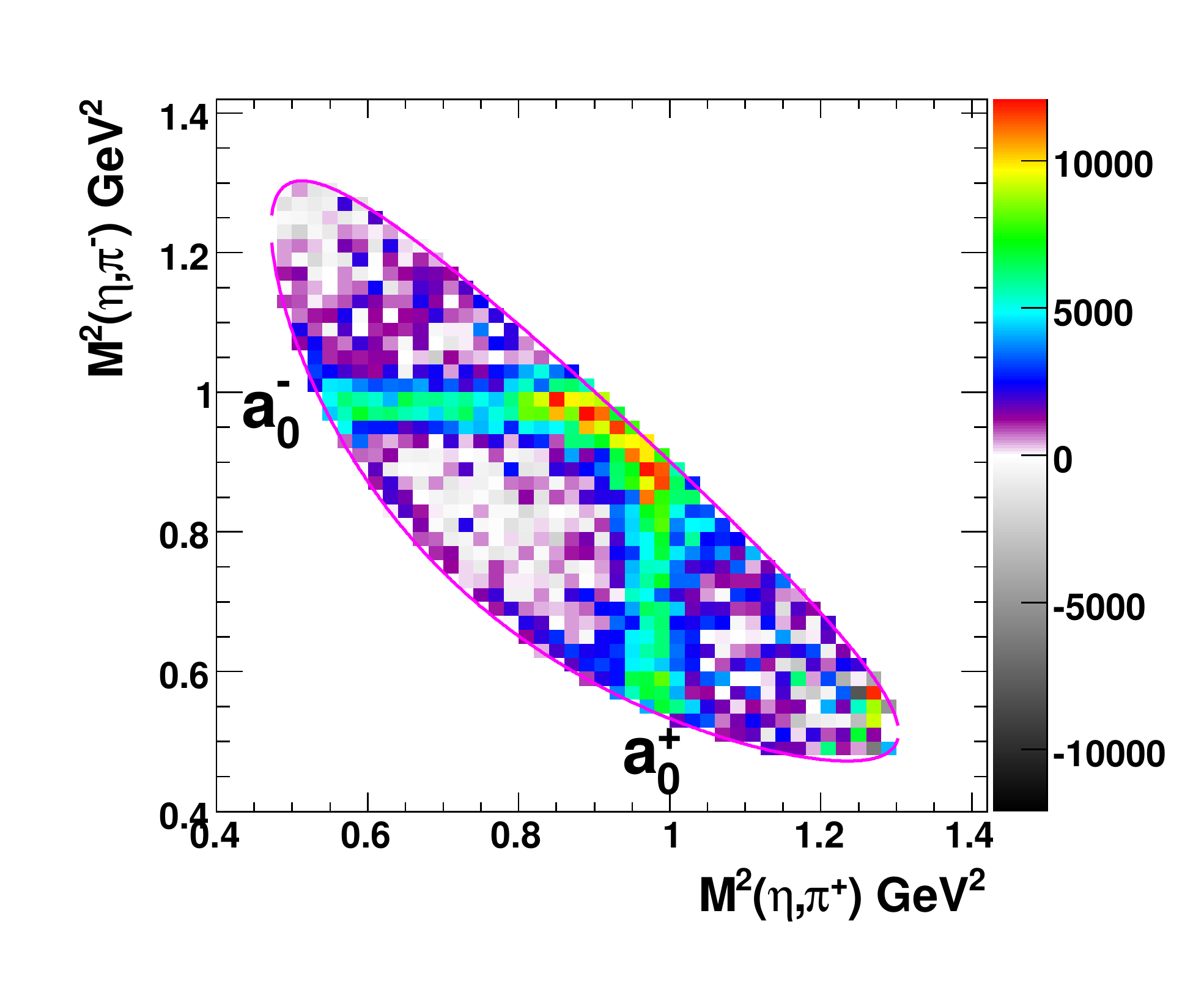}
  \caption{(Color online) Acceptance-corrected Dalitz plot for $\eta \pi^+ \pi^-$
    events with \mmp\ between 1251 and 1311 \mmunit\, after subtracting the weighted and scaled sidebands.
    }
  \label{fig:dalitz_acceptance_corrected_v6}
\end{figure}

The Dalitz plot for the $\eta\pi^+\pi^-$ decay after background subtraction and correction for acceptance is shown in Fig.~\ref{fig:dalitz_acceptance_corrected_v6}.  The lower right portion of the distribution has large bin-to-bin statistical fluctuations as expected in light of the low acceptance  of this kinematic region. There is a slight ``edge-effect'', an event excess along the edge of the allowed phase-space, that is  noticeable along the low-mass edge between the $a_0$ bands. This is due to imperfect sideband subtraction.  The main qualitative observation about the signal is that the $a_0^\pm(980)$ bands are of similar intensity.  There is a fairly thorough depletion of events between the two-body bands, and a lesser depletion toward the outer wings of the distribution. 

We tested the technique of sideband scaling using a ``toy'' Monte Carlo model, without \clas\ acceptance. Events were generated with a signal-to-background ratio approximating that seen in the $\gamma p \to p \pi^+ \pi^- (\eta)$ data. Both the \fx\ signal and background \etapipi\ events were generated according to 3-body phase space.  The sidebands of the \fx\ were scaled in the Dalitz mass variables, $m^2_{\eta\pi^+}$ and $m^2_{\eta\pi^-}$, according to the described technique. The sideband events were then subtracted from the central band of events having \mmp\ between 1251 and 1311 \mmunit.  This mass range was wide enough to produce a noticeable ``edge effect'' on the Dalitz plot due to imperfect mapping of the background kinematics onto the signal region.  Events lying outside the kinematic limit for a \mmp\ of 1281 \mmunit\  were removed.  No significant distortions in the resulting ``toy'' Dalitz plot were seen. The subtracted plot yielded about 11\% fewer events due to this trimming of the kinematic boundary.

We also examined the choice of mass range used to define the signal and sideband regions.  There is unavoidably some signal present in the sideband regions from the tails of the \xx\ meson. This leads, however, to a calculable over-subtraction of events. While it is possible to reduce this oversubtraction by widening the central signal region, this reduces the signal-to-background ratio and reduces the validity of our assumption that the kinematics of background events in the sidebands is similar to that in the \xx\ peak region. The dependence of our results on these choices was tested by varying the range of the mass bands.  We also tested introducing a gap between the signal band and sidebands.  The best result was found when choosing a 60~MeV-wide central region for the \xx, with 30~MeV-wide lower and upper sidebands, with no gap.

Finally, to look for any other biases in this procedure we ran our analysis on 10 million simulated $\gamma p \to p \rho^0 \pi \pi$ events. The Dalitz-plot  distribution for this background was not uniformly flat, but after using our method and correcting for acceptance, the resultant uniform Dalitz plot was statistically consistent with  zero false signal events. The consistency of both the toy-model tests and this background simulation leads us to conclude that the bands of the $a_0(980)\pi$ events are not significantly biased by our background-subtraction method.  A more detailed description of the sideband subtraction method is found in Ref.~\cite{Dickson:th}.


\subsection{Amplitude analysis of the decay distribution}
\label{sec:amplitude}

The strong $a_0^\pm(980)$ bands seen in Figure~\ref{fig:dalitz_acceptance_corrected_v6} show the decay of the parent state, be it the \fx, \etax, or both, occurs dominantly through the $a_0^\pm \pi^\mp$ intermediate states.  Furthermore, where the two bands nearly meet it appears that there is coherent addition of amplitudes, leading to considerable peaking, with a hint of additional peaking at the other ends of the bands. The $a_0$ and $\pi$ are spin zero states, so the spin of the parent \xx\ can be revealed in the relative orbital angular momentum between them.  If the \xx\ is the \etax, the decay products will be found in a spatial  $s$-wave configuration, while if the \xx\ is the \fx\ they will be found in $p$-wave.

In the case of a $p$-wave decay, the relevant quantization axis choice is important.  The decay angular distribution in the $L=1$ final state will have characteristic $m=0$ and $m=\pm1$ intensities with respect to the axis along which the spin-1 particle is aligned.  We tested the two usual cases:  the $s$-channel helicity (``helicity") system and the $t$-channel helicity (``Gottfried-Jackson") system.  

In the helicity system the quantization axis is that of the created meson in the overall reaction center-of-mass  frame.  If it is produced via the decay of an intermediate-state high-mass $N^*$ resonance, the $N^*$ and the final-state proton are then colinear in the meson rest frame.  The spin-1 meson is not required to be aligned along the $N^*$-$N$ axis, but if the reaction mechanism happens to create an alignment, it will be evident in the $a_0\pi$ angular distribution with respect to this axis.  In the Gottfried-Jackson system the quantization axis is the direction of the incoming photon in the rest frame of the produced meson.  This is the relevant axis if the particle is produced by, say, $\rho$  exchange in the $t$ channel, since in that system the photon and the exchanged $\rho$ are colinear.  Again, alignment is not required, but if it exists it will be seen in the angular distribution of the $a_0\pi$ decay products along the $\gamma$-$\rho$ axis.  The degree of alignment will be one of the results of the fitting procedure.   Depending on which angular distribution prevails in the decay process, different regions of the final $m_{\eta\pi^-}^2$ vs. $m_{\eta\pi^+}^2$ distribution, which is sensitive to the relative angles in the 3-particle final state, will be populated. Given the interference among the decay amplitudes, greater or lesser amounts of interference will be found at any given place in the Dalitz plot.

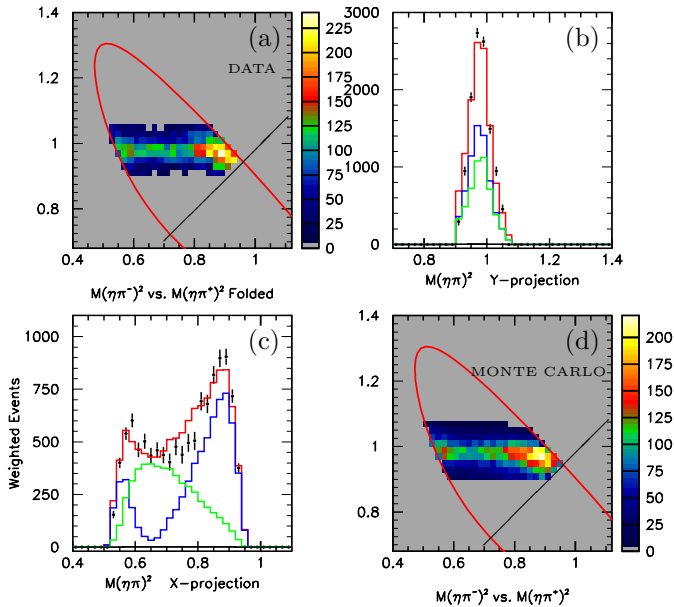
\begin{figure}[h!]
\centering
  \begin{overpic}[width=0.50\textwidth]{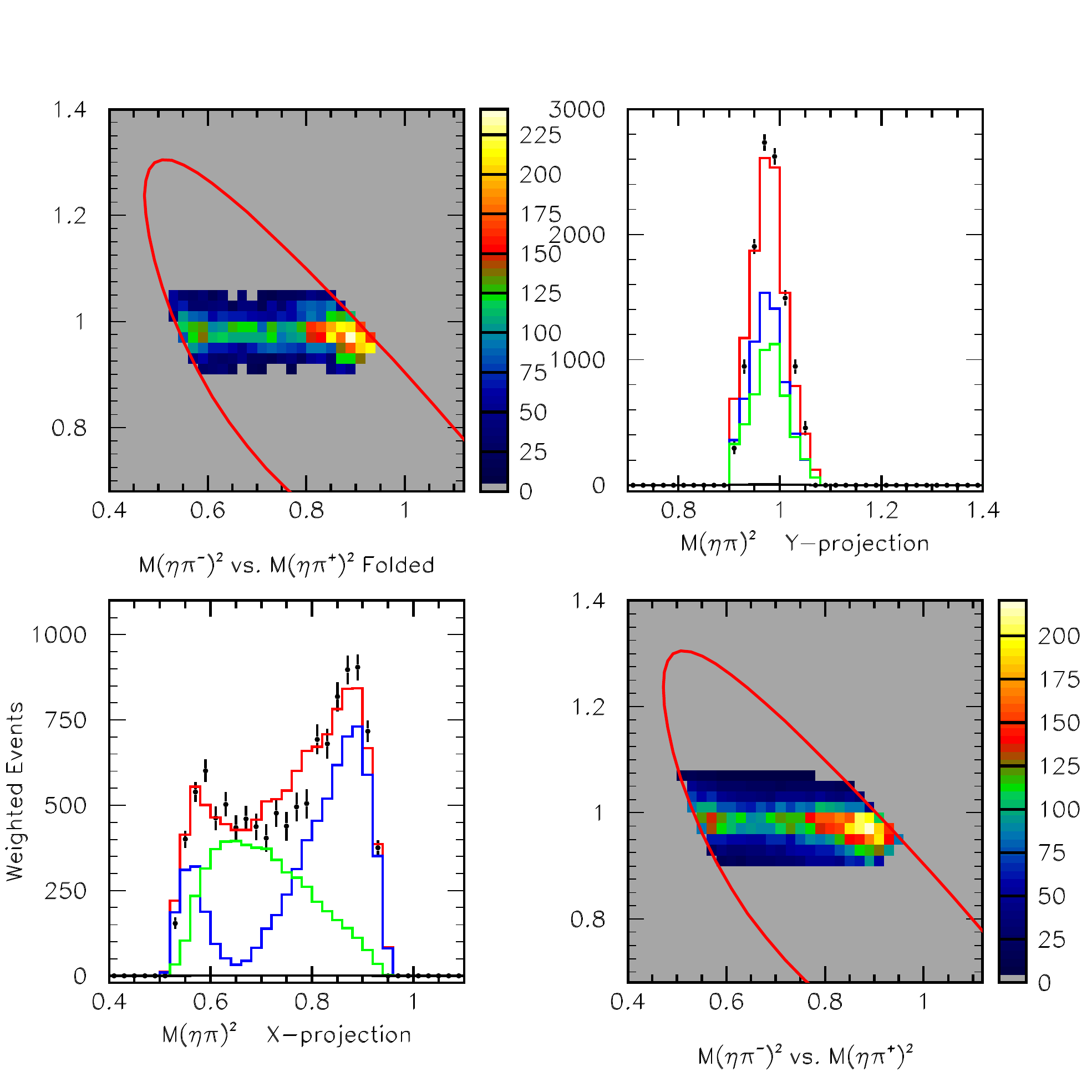}
    \put(36,85){(a)} 
    \put(83,85){(b)}
    \put(36,40){(c)}
    \put(83,40){(d)}
    \put(33,81){\tiny{DATA}} 
    \put(69,36){\tiny{MONTE CARLO}} 
    \put(23.5,56){\color{black}\line(1,1){18.5}  } 
    \put(71,11){\color{black}\line(1,1){18.5}  } 
  \end{overpic}
  \caption{(Color online) Helicity system fit to the $x \to \eta \pi^+ \pi^-$ Dalitz distribution.  Data  (a) and weighted Monte Carlo (d), as well as projections of both onto the vertical axis (b) and the horizontal axis (c).  For both data and Monte Carlo, events are restricted to the range of the dominant $a_0$ band of events.   Blue lines: coherent $L=0$ and 1,  $m=0$;  green line: $L=1$, $m=1$;  red line:  sum total.
  \label{fig:fig16}}
\end{figure}

The Dalitz-plot data was fitted starting with Monte Carlo events generated according to ``flat" phase space but with the measured width of the parent meson and the detector resolution function built in.  The events were trimmed to reside entirely inside the nominal boundary contour of decays using the mass centroid of the decaying state, as illustrated in Figs.~\ref{fig:uncorrected_dalitz_both} through \ref{fig:fig16}.

The decays $x \to a_0\pi$ in both charge states were modeled with the $a_0$ represented by a relativistic Breit-Wigner function with central mass $m_0$ and width $\Gamma_0$:
\begin{equation}
BW(m|m_0,\Gamma_0) = \frac{\sqrt{m_0\Gamma_0}}{m_0^2 - m^2 - i m_0 \Gamma_0 \frac{q(m)}{q(m_0)} }.
\end{equation}
Here $q(m)$ is the two-body break-up momentum of a parent state of mass $m_x$ to an $a_0$ of mass $m$ and a pion.  More formally, one writes $q = q(m,m_x,m_\pi)$ since the available breakup momentum depends upon all three masses.  We found that this relativistic Breit-Wigner form yielded results nearly identical to using the non-relativistic form with the ``$q(m)$" factors omitted, since the reaction kinematics is rather far from the decay thresholds.   The ratio $q(m)/q(m_0)$ was in the range 0.95 to 1.1.  The scalar $a_0$ particle has a complex structure~\cite{PhysRevD.59.012001} and could be described, for example, by a more accurate Flatt\'e-type parametrization~\cite{Flatte1976224}, but that was not needed for the present purpose.  

For each Monte Carlo event, both the $a_0^+\pi^-$ and $a_0^-\pi^+$ amplitudes were computed and added coherently.  That is, for each event the BW weight was computed using $m = m_{a_0^+ \pi^-}$, and then again for  $m = m_{a_0^- \pi^+}$ for the same event.   For the $L=0, m=0$ decay that characterizes the decay of an $\eta(1295)$ state, that is all that is needed. The decay is isotropic in the rest frame of the decaying state and it does not matter what quantization axis one chooses.  But for the $L=1, m=0,\pm 1$ decay of the $f_1(1285)$  state, the relevant angular correlations must be included.  Consider that photoproduction of the parent state produces, by some a priori unknown mechanism, a $J^P=1^+$  particle with a spin wave function
\begin{equation}
\chi_{f_1} = 
\left(
\begin{array}{c}
a\\
b\\
a\\
\end{array}
\right)
\label{eq:spinor}
\end{equation}
where $a$ and $b$ are the amplitudes for the $m=\pm 1$ and the $m=0$ substates, respectively.  We require 
\begin{equation}
b=\sqrt{1 - 2 a^2} 
\label{eq:N}
\end{equation}
for proper normalization.  The $p$-wave decay of this state into two spin-zero particles (the $a_0$ and the $\pi$) then leads to a spatial
wavefunction of the form 
\begin{equation}
W_{L=1, m=0,\pm 1}(\theta,\phi) = a Y_{1,+1}(\theta,\phi)+b Y_{1,0}(\theta,\phi) + a Y_{1,-1}(\theta,\phi),
\end{equation}
using the usual spherical harmonic functions, and where $\theta$ and $\phi$ are the decay angles in the $f_1(1285)$ rest frame with respect to the chosen coordinate system axes.   The parameter $a$ (and implicitly $b$) is determined in the fit.  The corresponding expression for decay into an $s$-wave final state from a $J^P = 0^-$ state is
\begin{equation}
W_{L=0, m=0}(\theta,\phi) = c Y_{0,0},
\end{equation}
where $c$ is introduced as another parameter of the fit.

The overall amplitude for the decay of the parent meson $x$ can then be expressed, for each Monte Carlo event, in terms of two amplitudes that do not interfere with each other by virtue of the orthonormality of the spherical harmonics.  The first is for the $m=\pm 1$ parts:
\begin{widetext}
\begin{equation}
A_{m=\pm 1}(m_{a_0^+ \pi^-}, m_{a_0^- \pi^+}) = BW(m_{a_0^+ \pi^-}) W_{1, \pm 1}(\theta_{a_0^+ \pi^-},\phi_{a_0^+ \pi^-})
+ BW(m_{a_0^- \pi^+}) W_{1, \pm 1}(\theta_{a_0^- \pi^+},\phi_{a_0^- \pi^+}). 
\end{equation}
\end{widetext}
The second is for the $m=0$ parts:
\begin{widetext}
\begin{equation}
A_{m=0}(m_{a_0^+ \pi^-}, m_{a_0^- \pi^+}) = BW(m_{a_0^+ \pi^-})
(W_{1, 0}(\theta_{a_0^+ \pi^-},\phi_{a_0^+ \pi^-}) + W_{0, 0})
+ BW(m_{a_0^- \pi^+})  
(W_{1, 0}(\theta_{a_0^- \pi^+},\phi_{a_0^- \pi^+}) + W_{0, 0}).
\end{equation}
\end{widetext}
The total magnitude-squared of the event, $T$,  is then computed and added to the relevant bin of the Dalitz plot according to 
\begin{widetext}
\begin{equation}
T(m_{a_0^+ \pi^-}, m_{a_0^- \pi^+})  = \frac{q(m_{a_0^+ \pi^-})}{q(m_0)}  \frac{q(m_{a_0^- \pi^+})}{q(m_0)}  \left( |A_{m=\pm 1}(m_{a_0^+ \pi^-}, m_{a_0^- \pi^+})|^2  + |A_{m=0}(m_{a_0^+ \pi^-}, m_{a_0^- \pi^+})|^2 \right).
\label{eq:T}
\end{equation}
\end{widetext}
The prefactors  represent the phase space for the final state of the event;  they are always close to unity.  This final expression does not exhibit the angles $\theta$ and $\phi$  at which the $a_0$ and  $\pi$ pair are created with respect to the chosen quantization axis.    The total magnitude-squared weight of each bin in the Dalitz plot is determined by a sum over all the Monte Carlo events generated in the simulation, sampling all possible polar and azimuthal angle combinations.  In this way the angular dependence of the decay, as it affects the Dalitz-plot distribution, is modeled by the calculation.  

The region outside the strong $a_0\pi$ bands was first modeled by a constant non-resonant amplitude.  However, this did not lead to satisfactory fit results:  there is broad structure (see Fig.~\ref{fig:dalitz_acceptance_corrected_v6}) that we were unable to describe.  Good fits were made by selecting only the events in the dominant bands between $m_{\eta\pi} = 0.95$~GeV and 1.03~GeV.   Fits were made according to both the helicity-system weighting of events and according to the Gottfried-Jackson system weighting.  By far the best result used the helicity system, which is shown in Fig.~\ref{fig:fig16}.   Figure~\ref{fig:fig16}a shows the data folded along the symmetry axis in the region that was used in the fit to the dominant decay amplitudes.  Figure~\ref{fig:fig16}b shows the projection of the data onto the vertical axis, emphasizing the  $a_0$ peak.  The data points are shown in black with statistical error bars.  The green histogram is for the  $L=1, m=\pm1$ contribution.  The blue histogram is for the combined interfering $L=1, m=0$ and $L=0, m=0$ component.  The statistical uncertainties associated with the weighted Monte Carlo are not shown because they are very small.  A non-resonant background made no significant contribution to this fit to the restricted data set, and so was not included in the fit.  The red histogram is the sum (non-interfering) of the components.  The mass of the scalar $a_0$ was in agreement with the PDG value, while the width was smaller at about 35~MeV.  Figure~\ref{fig:fig16}c shows the projection of the fit onto the horizontal axis.  The larger lobe at the upper end of the spectrum is a consequence of the interference between the $a_0$ bands.  Again, the blue histogram is for the interfering $L=1, m=0$ and $L=0, m=0$ resonant contributions.   The green histogram is the $L=1, m=\pm 1$ contribution.  Recall that the $m=0$ and $m=\pm 1$ components of the $L=1$ wave are not independent of each other.  There is only one fit parameter ($a$ in Eq.~\ref{eq:spinor}) that controls both.  The production mechanism of the spin-one $f_1$ state, while we do not know how it populates the different sub-states, must produce a coherent combination of $m=0$ and $m=\pm 1$, as codified in Eq.~\ref{eq:N}.    Figure~\ref{fig:fig16}d is the weighted Monte Carlo for visual comparison with the data in Fig.~\ref{fig:fig16}a.    

The same model applied in the Gottfried-Jackson system did not reproduce the large and small peaks seen in Fig.~\ref{fig:fig16}c.   Thus, it appears that the helicity system is the one preferred by the data.  According to this fit the overwhelming strength of the decay is in the $L=1$ component of the coherent sum, consistent with the decaying meson being the \fx\ state.  The $L=0$ strength consistent with a \etax\ was $0.06 \pm 0.01\%$, showing that essentially only the \fx\ is photoproduced in this reaction.  Furthermore, a fit allowing only the $L=0$ amplitude of the \etax\ also completed failed  to reproduce the coherent peak in the overlap region of the $a_0$ bands.  

We see that the fit in Fig.~\ref{fig:fig16}c is good but not perfect.  This may in part be due to the ``phase space" acceptance used in the  Dalitz plot simulation.  The event distribution was not  iteratively corrected to incorporate the fitted angular correlations in the data into the acceptance.

The fit leading to Fig.~\ref{fig:fig16} gives the fraction (or probability) of the parent meson state in the $L=1$ $m = \pm 1$ substates, $P_\pm$, and the fraction in the $m=0$ substate, $P_0$ (related to $a$ and $b$ in Eq.~\ref{eq:spinor}) .  The two portions add up to 100\%, by construction.  The proportion is
\begin{equation}
P_\pm : P_0 = 31.8 : 69.2,  \pm 1.4\%.
\label{eq:prop}
\end{equation}
That is, the reaction mechanism leading to formation of the $f_1(1285)$, integrated over all energies and angles, gives this proportion of the spin substates.   

Thus, we have evidence that the \fx\ is photoproduced dominantly via the decay of an excited $s$-channel  ($N^*$) system, and that its $J=1$ substates $m=\pm 1$ and  $m=0$ are populated in the reaction, averaged over all production angles and energies, in a ratio of roughly 1:2, as shown in Eq.~\ref{eq:prop}.  It is easy to show using Clebsch-Gordan algebra that the minimum spin of an $N^*$  decaying to this final state in $s$-wave is $J^P = (3/2)^+$.   If the decaying state had  spin $J^P =(1/2)^+$, the expected proportion would be reversed at 2:1.  There are, however, no known low-spin nucleon resonances in the mass range between 2.3 and 2.8 GeV.  The four-star $N(2220)$ has $J^P=(9/2)^+$, which would necessitate a decay with a minimum orbital angular momentum of $L=3$.  Thus, there are no candidate $N^*$ states that would allow for a simple explanation of this process.

As mentioned, the region in the Dalitz plot outside the dominant $a_0\pi$ decay bands was not consistent with zero, as seen in  Fig.~\ref{fig:dalitz_acceptance_corrected_v6}.  With the present statistics we see no clear structure, but the distribution is not uniform, either.  The Particle Data Group lists, as one of the $f_1(1285)$ branching ratios, the fraction of decay to $a_0\pi$ (ignoring decays to  $K\bar{K}$) to the decay of the $f_1(1285)$ to any $\eta\pi\pi$ final state.  In the PDG notation this is called  ``$\Gamma_9/\Gamma_8$".   We estimated this ratio from the present experiment.  Since we do not know the reaction mechanism leading to decay outside of the strong bands, we proceeded as follows.

\begin{enumerate}
\item
Select suitable bands defining the region of the $a_0$ decays.   Sum the bins within the bands, which represents the strength of the dominant $a_0\pi$ decay with some contamination from the other decay mechanism of unknown nature.  Total ``counts" in this range are called $N_B$.  

\item
Sum the bins outside the bands, which represent the sub-dominant decays.  Define total ``other counts" as  $N_O$.  The sum of everything in the whole Dalitz distribution is $N_B+N_O$.

\item
Reduce $N_B$ by the estimated amount of ``flat" sub-dominant decay underneath the bands using the fractional area of the banded region to the whole plot region.  This area ratio was about $r_c=0.55$ for the optimal band selection.  We ignore any interference of the dominant and the sub-dominant decay mechanisms.

\item
Compute the desired branching ratio using
\begin{equation}
\frac{\Gamma( a_0\pi \text{ (no } K\bar{K}))}{\Gamma(\eta\pi\pi\text{ (total)})} = \frac{N_B - N_O r_c}{N_B + N_O}
\end{equation}

\item
Compute the statistical uncertainty from the two independent measurements.  This was 2\%.  The systematic uncertainty was estimated.  The correction factor in the numerator of the expression is certainly not zero, but the assumption that the non-dominant decay is ``flat'' is also not accurate.  Thus, we estimated that the systematic uncertainty is as large as the correction itself, that is, of size $N_O r_c$.  The Gaussian estimator of a quantity that has a uniform probability density in some range $A$ is $A/\sqrt{12}$.  Thus, the systematic uncertainty from this source is $N_O r_c/\sqrt{12}$.  This was about 3\%.

\item
We varied the width of the band that defines the $a_0$ region in the Dalitz plot over a plausible range.  This changed the division between the dominant and the non-dominant decay mechanisms.  We estimated the systematic uncertainty due to this source as  about 9\%.

\end{enumerate}
With the above considerations, we estimate the branching ratio to be 
\begin{equation}
\frac{\Gamma( a_0\pi \text{ (no } K\bar{K}))}{\Gamma(\eta\pi\pi\text{ (total)})} = 74 \pm 2 (stat) \pm 9 (syst) \%
\label{eq:other}
\end{equation}
Without a comprehensive theoretical model for the decay of the $f_1(1295)$  no more precise estimate was possible. For comparison, the present PDG value is $69 \pm 13\%$.  Thus, the present result is consistent with the world average.

\section{Discussion and Conclusions}
\label{sec:conclusions}

Using the CLAS system we have investigated, for the first time, properties of the narrow meson seen  in photoproduction from the proton at a mass  $m_0=1281.0 \pm 0.8$~MeV and with width $\Gamma = 18.4 \pm 1.4$~MeV.  The measured mass and width are  more compatible with the known properties~\cite{Agashe:2014kda} of the \fx\ than the \etax.  The measured width is, however, about 6 MeV smaller than the previous world average. This may be due to our careful removal of the intrinsic experimental resolution, leaving only the Breit-Wigner component of the width.

The highest statistics were found in the $\eta\pi^+\pi^-$ decay mode, but the meson was also reconstructed from the $\bar{K}K\pi$ and $\gamma\rho^0$ modes.  No evidence was found for any of the higher mass $0^{-+}$ or $1^{++}$ states $\eta(1405)$, $\eta(1470)$, $f_1(1420)$ ßor $f_1(1510)$ in these decay modes.

The cross section is much ``flatter" in angle than that of the nearby $\eta^\prime(958)$.  Comparison of the differential cross sections with meson-exchange model predictions show more strength at central and backward angles than achievable through only $t$-channel production processes. Only at forward angles and higher energy bins does the Kochelev model~\cite{Kochelev:PhysRevC.80.025201} approach the data  in magnitude. The other model predictions gave poorer agreement with experimental data.  This suggests that the production mechanism is not mainly $t$-channel.

The observed branching ratio ${\Gamma(K\bar{K}\pi)}/{\Gamma(\eta\pi\pi)} = 0.216 \pm 0.032$ is consistent with the PDG value of $0.171 \pm 0.013$ for the \fx. There is no world data for this ratio for the \etax, though it has been observed in $K\bar{K}\pi$ final states with strength comparable to the \fx~\cite{Adams:2001sk}.

The radiative decay branching ratio  ${\Gamma(\gamma\rho^0)}/\Gamma({\eta\pi\pi)}$ is found to be $0.047 \pm 0.018$, which is less than half the PDG average value of $0.105 \pm 0.022$,  inconsistent  by about 3 standard deviations.  Nevertheless, the presence of the signal in the $\gamma\rho^0$ decay mode supports the \fx\ identity of the observed state, as seen from spin and parity considerations.  The axial-vector \fx\ can couple to \rhogamma\ via the $E1$ multipole, while the pseudoscalar \etax\ can couple only via $M1$.  In the $t$-channel the \fx\ should be dominant in photoproduction.  This argument is weakened, however, by the observation that $t$-channel may not be the dominant photoproduction channel for this meson; the experimental results show there may be more coming from $s$-channel processes.
 
The Dalitz distribution of  the $\etapipi$  final state shows that this decay occurs primarily through an $a_0^\pm \pi^\mp$ intermediate state, with the $a_0^\pm$ subsequently decaying to $\eta \pi^\pm$.   Other decay mechanisms may account for about a quarter of the total (Eq.~\ref{eq:other}).  There is constructive interference between the $a_0$ bands, and amplitude analysis shows this can be reproduced with amplitudes written in the $s$-channel helicity system.  The decay of the parent meson to $a_0\pi$ is overwhelmingly  in $p$-wave, indicating that the meson has quantum numbers $J^P=1^+$, proving it to be the \fx.  The alignment of the \fx\, averaged over the kinematics of this measurement, was measured by fitting the decay angular distributions (Eq.~\ref{eq:prop}).  If the state were produced by the decay of a low-spin $N^*$ state, the baryon would have $J^P=(3/2)^+$,  but there are no such candidate states in the PDG listings.   Any interfering $0^-$ wave indicative of excitation of an \etax\ is at a vanishingly-small sub-percent level.  The Dalitz distribution is not reproduced with amplitudes computed in the Gottfried-Jackson system.  This supports the conclusion that the \fx\ is photoproduced via an $s$-channel process, involving an $N^*$ excitation or a process related to the possible $\bar{K}K^*$ molecular nature of the $f_1(1285)$.   

Taken together, the results from the suite of measurements in this analysis support the conclusion that the meson state observed in \clas\ photoproduction is the well-known  $J^{PC}=1^{++}$ \fx.  The interference of the dominant $a_0^\pm \pi^\mp$ bands in the Dalitz distribution, the presence of radiative decays to $\gamma \rho^0$, and the measured mass that is consistent with world data support this identification.  The smaller measured intrinsic width, and the smaller radiative branching ratio of $\gamma\rho^0$ to $\eta\pi\pi$  are not enough to spoil this conclusion.

There is disagreement between the \fx\ cross section and predictions by $t$-channel based photoproduction models.  It has an angular distribution less steep than  other meson photoproduction channels, and there is dominance of $\eta\pi\pi$ decays in the $s$-channel helicity system rather than the Gottfried-Jackson system.  These findings suggest that the dynamical nature of this state and its photoproduction are not yet understood, but may be found in an $s$-channel production mechanism.

\begin{acknowledgments}
We acknowledge the outstanding efforts of the staff of the Accelerator
and Physics Divisions at Jefferson Lab that made this experiment
possible. The work of the Quark Interactions group at Carnegie
Mellon University was supported by DOE grant DE-FG02-87ER40315.  The
Southeastern Universities Research Association (SURA) operated the
Thomas Jefferson National Accelerator Facility for the United States
Department of Energy under contract DE-AC05-84ER40150.  Further
support was provided by the National Science Foundation, the United
Kingdom's Science and Technology Facilities Council grant ST/J000175, and  the Italian 
Istituto Nazionale di Fisica Nucleare.
\end{acknowledgments}

%
%


\appendix
\section{Partial \fx\ Cross Section to the \etapipi\ Final State}
\label{apx:dcs_numbers}

The differential cross sections in Table~\ref{tab:table_diff} are the weighted mean of independent measurements in the \etapipi\ and \kkpi\ decay modes of the \fx. The \kkpi\ results were scaled using the present measurement of the branching ratio $\Gamma(\fx~\to~\kkpi) / \Gamma(\fx \to \etapipi)$. This was done for improved statistical precision.  The results have \emph{not} been corrected for the unmeasured (by us) branching fraction $\Gamma(\fx \to \etapipi) / \Gamma_{total}$. The given systematic uncertainty  $\sigma_{sys}$ includes the sources listed in Table~\ref{tab:final_systematics} and those discussed in Sec.~\ref{sec:results}.  Electronic tabulations of the results are available from several sources: Refs.~\cite{clasdb}, \cite{durham}, \cite{contact}.

\vspace{1.0cm}

\begin{longtable}{ccccc}

  \caption{Differential cross section for $\gamma p \to \fx p \to \eta  \pi^+ \pi^- p$ in nanobarns/steradian. The point-to-point uncertainties are given in
    separate statistical and systematic contributions.
  \label{tab:table_diff}
  }\\
\hline\hline 
$W    $  & $\cos\Theta^{c.m.}$ & $\frac{d\sigma}{d\Omega}$ & $\sigma_{stat}$ &  $\sigma_{sys}$ \\
(GeV)    &                     & (nb/sr)                   & (nb/sr)         &  (nb/sr)        \\
\hline
\endfirsthead

$W    $  & $\cos\Theta^{c.m.}$ & $\frac{d\sigma}{d\Omega}$ & $\sigma_{stat}$ &  $\sigma_{sys}$ \\
(GeV)    &                     & (nb/sr)                   & (nb/sr)         &  (nb/sr)        \\
\hline
\endhead

\hline \hline
\endlastfoot
& \\[0.1em]
2.35 & -0.70 & 5.96 & 0.57 & 1.57 \\[1 pt]
2.35 & -0.50 & 4.70 & 0.45 & 0.66 \\[1 pt] 
2.35 & -0.30 & 6.42 & 0.54 & 0.90 \\[1 pt]
2.35 & -0.10 & 8.37 & 0.74 & 1.26 \\[1 pt]
2.35 & 0.10 & 8.29 & 0.67 & 1.17 \\ [1 pt]
2.35 & 0.30 & 7.81 & 0.64 & 1.26 \\ [1 pt]
2.35 & 0.50 & 7.42 & 0.76 & 1.05 \\ [1 pt]
2.35 & 0.70 & 5.01 & 0.58 & 0.70 \\ [1 pt]
2.35 & 0.85 & 3.18 & 1.00 & 0.45 \\ [1 pt]
& \\[0.1em]
2.45 & -0.70 & 2.80 & 0.40 & 0.93 \\ [1 pt]
2.45 & -0.50 & 6.42 & 0.60 & 1.07 \\ [1 pt]
2.45 & -0.30 & 6.39 & 0.63 & 1.32 \\ [1 pt]
2.45 & -0.10 & 5.73 & 0.48 & 0.75 \\ [1 pt]
2.45 & 0.10 & 7.29 & 0.52 & 1.04 \\  [1 pt]
2.45 & 0.30 & 8.10 & 0.60 & 1.96 \\  [1 pt]
2.45 & 0.50 & 6.58 & 0.62 & 1.16 \\  [1 pt]
2.45 & 0.70 & 6.68 & 0.59 & 1.45 \\  [1 pt]
2.45 & 0.85 & 2.16 & 1.05 & 0.26 \\  [1 pt]
& \\[0.1em]
2.55 & -0.70 & 4.22 & 0.49 & 0.88 \\  [1 pt]
2.55 & -0.50 & 3.50 & 0.30 & 0.44 \\  [1 pt]
2.55 & -0.30 & 4.38 & 0.51 & 1.46 \\  [1 pt]
2.55 & -0.10 & 5.37 & 0.42 & 1.10 \\  [1 pt]
2.55 & 0.10 & 6.57 & 0.48 & 1.08 \\  [1 pt]
2.55 & 0.30 & 6.70 & 0.52 & 1.01 \\  [1 pt]
2.55 & 0.50 & 12.12 & 0.78 & 2.66 \\  [1 pt]
2.55 & 0.70 & 9.70 & 1.04 & 1.46 \\  [1 pt]
2.55 & 0.85 & 7.95 & 1.10 & 2.20 \\  [1 pt]
& \\[0.1em]
2.65 & -0.70 & 2.21 & 0.35 & 0.37 \\  [1 pt]
2.65 & -0.50 & 2.58 & 0.33 & 0.39 \\  [1 pt]
2.65 & -0.30 & 2.85 & 0.37 & 0.40 \\  [1 pt]
2.65 & -0.10 & 3.73 & 0.31 & 0.84 \\  [1 pt]
2.65 & 0.10 & 3.03 & 0.28 & 0.86 \\  [1 pt]
2.65 & 0.30 & 4.26 & 0.43 & 0.86 \\  [1 pt]
2.65 & 0.50 & 8.40 & 0.49 & 1.40 \\  [1 pt]
2.65 & 0.70 & 9.81 & 1.06 & 2.40 \\  [1 pt]
2.65 & 0.85 & 0.50 & 1.06 & 1.90 \\  [1 pt]
& \\[0.1em]
2.75 & -0.70 & 2.49 & 0.28 & 0.32 \\  [1 pt]
2.75 & -0.50 & 1.55 & 0.26 & 0.20 \\  [1 pt]
2.75 & -0.30 & 1.70 & 0.18 & 0.30 \\  [1 pt]
2.75 & -0.10 & 1.71 & 0.28 & 0.44 \\  [1 pt]
2.75 & 0.10 & 1.95 & 0.19 & 0.25 \\  [1 pt]
2.75 & 0.30 & 4.01 & 0.41 & 1.00 \\  [1 pt]
2.75 & 0.50 & 5.15 & 0.45 & 1.82 \\  [1 pt]
2.75 & 0.70 & 9.26 & 0.90 & 1.29 \\  [1 pt]
2.75 & 0.85 & 5.73 & 1.21 & 0.74 \\  [1 pt]

\end{longtable}


\section{\etapr\ Cross Section from the  \etapipi\ and \rhogamma\ Final States}
\label{apx:dcs_numbers_etap}

Results for \etapr\ shown in this paper (Fig.~\ref{fig:R_dcs_etaprime_compare}) are given in Table~\ref{tab:table_etap_diff}, for comparison to previous results using the same data set.  The listed deviation from previous CLAS published results~\cite{Williams:2009yj} (using a different analysis method) used a cubic spline interpolation between points in $\cos\Theta^{c.m.}$  in order to compute differences from present values. These differences were included in the systematic uncertainty estimation for the present results for the \fx.

\renewcommand*{\arraystretch}{0.7}
\begin{longtable}{ccccc}

\caption{Differential cross section for $\gamma p \to \eta^{\prime} p$  in nanobarns/steradian. \label{tab:table_etap_diff} }\\
\hline\hline 
$W $  & $\cos \Theta^{c.m.}$ & $\frac{d\sigma}{d\Omega}$ &$\sigma_{stat}$ & Deviation\\ 
(GeV) &                      & (nb/sr)                   & (nb/sr)        & (\%) \\
\hline
\endfirsthead

$W    $  & $\cos \Theta^{c.m.}$ & $\frac{d\sigma}{d\Omega}$ &$\sigma_{stat}$  & Deviation\\
(GeV)    &                      & (nb/sr)                   & (nb/sr)         & (\%) \\
\hline
\endhead

\hline \hline
\endlastfoot
& \\[0.1em]
2.05 & -0.70 & 38.92 & 1.41 & -5.8 \\  [1 pt]
2.05 & -0.50 & 44.73 & 1.56 & -4.6 \\  [1 pt]
2.05 & -0.30 & 49.33 & 1.69 & -4.7 \\  [1 pt]
2.05 & -0.10 & 54.21 & 1.79 & -14.7 \\  [1 pt]
2.05 & 0.10 & 58.02 & 1.91 & -5.7 \\  [1 pt]
2.05 & 0.30 & 65.76 & 2.12 & -7.4 \\  [1 pt]
2.05 & 0.50 & 71.30 & 2.32 & -12.9 \\  [1 pt]
2.05 & 0.70 & 77.96 & 2.60 & -1.9 \\  [1 pt]
2.05 & 0.85 & 68.89 & 2.42 & -9.8 \\  [1 pt]
& \\[0.1em]
2.15 & -0.70 & 24.08 & 0.94 & -4.1 \\  [1 pt]
2.15 & -0.50 & 24.82 & 0.93 & -8.3 \\  [1 pt]
2.15 & -0.30 & 25.45 & 0.92 & -4.6 \\  [1 pt]
2.15 & -0.10 & 29.00 & 1.03 & -8.8 \\  [1 pt]
2.15 & 0.10 & 33.80 & 1.19 & -5.6 \\  [1 pt]
2.15 & 0.30 & 48.23 & 1.66 & -5.0 \\  [1 pt]
2.15 & 0.50 & 70.21 & 2.30 & -7.9 \\  [1 pt]
2.15 & 0.70 & 98.44 & 3.21 & -7.1 \\  [1 pt]
2.15 & 0.85 & 80.09 & 3.36 & -25.0 \\  [1 pt]
& \\[0.1em]
2.25 & -0.70 & 20.72 & 0.87 & -9.3 \\  [1 pt]
2.25 & -0.50 & 14.78 & 0.64 & -8.9 \\  [1 pt]
2.25 & -0.30 & 9.68 & 0.41 & -1.5 \\  [1 pt]
2.25 & -0.10 & 9.91 & 0.45 & -7.9 \\  [1 pt]
2.25 & 0.10 & 13.07 & 0.56 & 4.2 \\  [1 pt]
2.25 & 0.30 & 22.84 & 0.86 & -1.4 \\  [1 pt]
2.25 & 0.50 & 44.01 & 1.55 & -6.5 \\  [1 pt]
2.25 & 0.70 & 74.04 & 2.56 & -11.3 \\  [1 pt]
2.25 & 0.85 & 81.93 & 3.48 & -11.8 \\  [1 pt]
& \\[0.1em]
2.35 & -0.70 & 15.14 & 0.63 & -7.4 \\  [1 pt]
2.35 & -0.50 & 8.51 & 0.48 & 2.1 \\  [1 pt]
2.35 & -0.30 & 4.85 & 0.29 & -18.4 \\  [1 pt]
2.35 & -0.10 & 6.27 & 0.38 & -5.5 \\  [1 pt]
2.35 & 0.10 & 9.14 & 0.44 & -20.8 \\  [1 pt]
2.35 & 0.30 & 15.17 & 0.61 & -10.4 \\  [1 pt]
2.35 & 0.50 & 30.77 & 1.23 & -5.4 \\  [1 pt]
2.35 & 0.70 & 57.36 & 2.16 & -11.3 \\  [1 pt]
2.35 & 0.85 & 75.88 & 3.34 & -8.9 \\  [1 pt]
& \\[0.1em]
2.45 & -0.70 & 8.31 & 0.40 & -0.8 \\  [1 pt]
2.45 & -0.50 & 3.20 & 0.22 & 6.4 \\  [1 pt]
2.45 & -0.30 & 2.62 & 0.18 & 7.0 \\  [1 pt]
2.45 & -0.10 & 4.17 & 0.26 & -3.9 \\  [1 pt]
2.45 & 0.10 & 7.21 & 0.34 & 3.2 \\  [1 pt]
2.45 & 0.30 & 10.11 & 0.45 & -2.4 \\  [1 pt]
2.45 & 0.50 & 21.41 & 0.89 & -3.1 \\  [1 pt]
2.45 & 0.70 & 46.14 & 1.86 & -6.1 \\  [1 pt]
2.45 & 0.85 & 67.09 & 3.07 & -31.2 \\  [1 pt]
& \\[0.1em]
2.55 & -0.70 & 4.70 & 0.26 & -5.5 \\  [1 pt]
2.55 & -0.50 & 1.44 & 0.17 & 0.1 \\  [1 pt]
2.55 & -0.30 & 2.06 & 0.14 & 11.5 \\  [1 pt]
2.55 & -0.10 & 3.95 & 0.23 & -1.4 \\  [1 pt]
2.55 & 0.10 & 4.79 & 0.25 & -6.6 \\  [1 pt]
2.55 & 0.30 & 5.96 & 0.32 & 4.9 \\  [1 pt]
2.55 & 0.50 & 11.93 & 0.63 & -4.6 \\  [1 pt]
2.55 & 0.70 & 37.34 & 1.68 & -9.1 \\  [1 pt]
2.55 & 0.85 & 63.72 & 3.16 & -21.8 \\  [1 pt]
& \\[0.1em]
2.65 & -0.70 & 2.17 & 0.15 & -7.4 \\  [1 pt]
2.65 & -0.50 & 0.81 & 0.08 & 3.3 \\  [1 pt]
2.65 & -0.30 & 1.64 & 0.11 & -6.0 \\  [1 pt]
2.65 & -0.10 & 2.78 & 0.14 & 2.0 \\  [1 pt]
2.65 & 0.10 & 3.05 & 0.18 & -16.7 \\  [1 pt]
2.65 & 0.30 & 2.41 & 0.17 & 8.4 \\  [1 pt]
2.65 & 0.50 & 5.94 & 0.39 & 1.9 \\  [1 pt]
2.65 & 0.70 & 24.10 & 1.24 & -2.7 \\  [1 pt]
2.65 & 0.85 & 53.44 & 2.82 & -32.5 \\  [1 pt]
& \\[0.1em]
2.75 & -0.70 & 1.14 & 0.10 & -43.5 \\  [1 pt]
2.75 & -0.50 & 0.56 & 0.07 & -7.9 \\  [1 pt]
2.75 & -0.30 & 1.13 & 0.08 & -17.0 \\  [1 pt]
2.75 & -0.10 & 1.47 & 0.09 & -17.0 \\  [1 pt]
2.75 & 0.10 & 1.62 & 0.13 & -25.5 \\  [1 pt]
2.75 & 0.30 & 1.43 & 0.17 & -24.4 \\  [1 pt]
2.75 & 0.50 & 3.35 & 0.32 & -6.8 \\  [1 pt]
2.75 & 0.70 & 16.89 & 1.09 & -17.3 \\  [1 pt]
2.75 & 0.85 & 40.43 & 2.96 & -2.4 \\  [1 pt]
\end{longtable}

\bibliography{CLAS_f1_photoproduction}


\end{document}